\documentclass[usenatbib,usegraphicx]{mn2e}
\usepackage{xspace}
\usepackage{amsmath}
\usepackage{amssymb}
\renewcommand{\d}{\partial}
\newcommand{\Nbody}{\textsl{N}-body\xspace}
\newcommand{\SMILE}{\textsl{SMILE}\xspace}
\newcommand{\lmax}{l_\mathrm{max}}
\newcommand{\nmax}{n_\mathrm{rad}}
\newcommand{\ngrid}{n_\mathrm{grid}}
\newcommand{\tdyn}{T_\mathrm{dyn}}

\title[New code for orbit analysis and Schwarzschild modelling]
{A new code for orbit analysis and Schwarzschild modelling of triaxial stellar systems}

\author[E. Vasiliev]{Eugene Vasiliev$^{1,2,3}$\thanks{E-mail: eugvas@lpi.ru}\\
$^{1}$Lebedev Physical Institute, Leninsky prospekt 53, Moscow, Russia\\
$^{2}$Rochester Institute of Technology, 76 Lomb Memorial dr., Rochester, NY 14623, USA\\
$^{3}$Laboratoire d'Astrophysique de Marseille, 38 rue Joliot Curie, 13388 Marseille, France}

\begin{document}

\date{July 1, 2013}

\pagerange{\pageref{firstpage}--\pageref{lastpage}} \pubyear{2013}

\maketitle

\label{firstpage}

\begin{abstract}
We review the methods used to study the orbital structure and chaotic properties of various 
galactic models and to construct self-consistent equilibrium solutions by the 
Schwarzschild's orbit superposition technique. 
These methods are implemented in a new publicly available software tool, \SMILE, 
which is intended to be a convenient and interactive instrument for studying a variety of 
2D and 3D models, including arbitrary potentials represented by a basis-set expansion, 
a spherical-harmonic expansion with coefficients being smooth functions of radius (splines), 
or a set of fixed point masses.
We also propose two new variants of Schwarzschild modelling, in which the density of each orbit 
is represented by the coefficients of the basis-set or spline spherical-harmonic expansion, 
and the orbit weights are assigned in such a way as to reproduce the coefficients of the 
underlying density model.
We explore the accuracy of these general-purpose potential expansions and show that 
they may be efficiently used to approximate a wide range of analytic density models and 
serve as smooth representations of discrete particle sets (e.g.\ snapshots from an \Nbody 
simulation), for instance, for the purpose of orbit analysis of the snapshot.
For the variants of Schwarzschild modelling, we use two test cases -- a triaxial Dehnen model 
containing a central black hole, and a model re-created from an \Nbody snapshot obtained by 
a cold collapse. 
These tests demonstrate that all modelling approaches are capable of creating equilibrium models.
\end{abstract}


\section{Introduction}

The study of galactic structure relies heavily on construction and analysis of self-consistent 
stationary models, in which stars and other mass components (i.e.\ dark matter) move in the 
gravitational potential $\Phi$, related to their density distribution $\rho$ by the Poisson 
equation
\begin{equation}  \label{eq_poisson}
\nabla^2 \Phi(\boldsymbol{r}) = 4\pi\,\rho(\boldsymbol{r}) \;,
\end{equation}
in such a way that the density profile remains unchanged.
The evolution of distribution function $f(\boldsymbol{r}, \boldsymbol{v})$ of stars moving 
in the smooth potential is described by the collisionless Boltzmann equation:
\begin{equation}  \label{eq_CBE}
\frac{\d f}{\d t} + \boldsymbol{v}\,\frac{\d f}{\d\boldsymbol{r}} - 
  \frac{\d\Phi}{\d\boldsymbol{r}}\, \frac{\d f}{\d\boldsymbol{v}} = 0 .
\end{equation}
To construct a stationary self-consistent model corresponding to a given density profile 
$\rho(\boldsymbol{r})$, one needs to find a function $f$ which is the solution of 
(\ref{eq_CBE}) with a potential $\Phi$ related to $\rho$ via (\ref{eq_poisson}), such that
$\rho=\int f(\boldsymbol{r}, \boldsymbol{v}) \, d\boldsymbol{v}$.
Various approaches to this problem may be broadly classified into methods based on the 
distribution function, Jeans equations, orbit superposition, and iterative \Nbody 
schemes (see introduction sections in \citet{Dehnen09, MorgantiGerhard12} for a nice summary).
Of these, the first two are dealing with a smooth description of the system, but are usually 
restricted to sufficiently symmetric potentials (or special cases like a triaxial fully 
integrable model of \citet{vdVenHVZ03}), 
while the latter two techniques represent the stellar system in terms of Monte-Carlo sampling 
of the distribution function by orbits or \Nbody particles, thus achieving more flexibility 
at the expense of lack of smoothness. 

In this paper, we discuss the Schwarzschild's orbit superposition method, first introduced 
by Martin \citet{Schw79}. 
It consists of two steps: first a large number of trajectories are computed numerically in 
the given potential $\Phi$ and their properties (most importantly, spatial shape) are stored 
in some way, then these orbits are assigned non-negative weights so that the density $\rho$, 
which corresponds to this potential via the Poisson equation (\ref{eq_poisson}), is given by 
a weighted sum of densities of individual orbits. 
In the classical Schwarzschild method, the density is represented in a discrete way by dividing 
the configuration space into a 3D grid and computing the mass in each grid cell, both for the 
underlying density profile to be modelled and for the fraction of time that each orbit spends 
in each cell. Then the contribution of each orbit to the model is obtained by solving a linear 
system of equations, subject to the condition that all orbit weights are non-negative. 
We propose two additional formulations of the Schwarzschild method, in which the density is 
represented not on a grid, but as coefficients of expansion over some basis.

This method was first applied to demonstrate that a triaxial self-consistent model of a galaxy 
can be constructed, and has been used to study various triaxial systems with central density cores 
\citep{Statler87} and cusps \citep{MerrittFridman96, Merritt97, Siopis99, Terzic02, Thakur07, 
Capuzzo07}. The essential questions considered in these and other papers are whether a particular 
potential-density pair can be constructed by orbit superposition, what are the restrictions on 
the possible shapes of these models, how important are different types of orbits, including 
the role of chaotic orbits.

The Schwarzschild method is also very instrumental in constructing mass models of individual 
galaxies. In this application the density model is obtained from observational data of surface 
brightness profile, which, however, doesn't have an unique deprojection in a non-spherical case.
Kinematic constraints also come from observations. Typically one constructs a series of 
models with varying shape, mass of the central black hole, etc., and evaluates the goodness of 
fit to the set of observables; the $\chi^2$ statistics is then used to find the range of possible 
values for the model parameters allowed by the observations. 
Due to complicated geometry of the triaxial case, involving two viewing angles, most of the 
studies concentrated on the axisymmetric models. There exist several independent implementations
of the axisymmetric modelling technique \citep[e.g.][]{CrettonZMR99, Gebhardt00, ValluriME04} 
and at least one for triaxial models \citep{vdBosch08}. Observation-driven Schwarzschild method 
was used for constructing models of the Galactic bulge \citep{Zhao96b, Hafner00, WangZMR12}, 
constraining mass-to-light ratio and shape of galaxies \citep{Thomas04, vdBosch08}, 
and for estimating the masses of central black holes \citep{Gebhardt03, Valluri05, vdBoschZeeuw10}. 
Related techniques are the made-to-measure method \citep{SyerTremaine96, deLorenzi07, Dehnen09, 
LongMao10} or the iterative method \citep{Rodionov09}, in which an \Nbody representation of 
a system is evolved in such a way as to drive its properties towards the required values, 
by adjusting masses or velocities of particles  ``on-the-fly'' in the course of simulation.

Of these two flavours of the Schwarzschild method, this paper deals with the first. 
We continue and extend previous theoretical studies of triaxial galactic models with arbitrary 
density profiles and, possibly, a central massive black hole.  
For these potentials which support only one classical integral of motion -- the energy%
\footnote{Axisymmetric Schwarzschild models are often called three-integral models, 
to distinguish them from simpler approaches involving only two classical integrals; 
however this is not quite correct since not all orbits respect three integrals of motion 
even in the axisymmetric case.} 
(in the time-independent case) -- orbital structure is often quite complicated and has various 
classes of regular and many chaotic orbits, therefore it is necessary to have efficient 
methods of orbit classification and quantifying the chaotic properties of individual orbits
and the entire potential model. 
These orbit analysis methods are often useful by themselves, besides construction of 
self-consistent models, for instance, for the purpose of analyzing the mechanisms driving 
the evolution of shape in \Nbody simulations \citep{Valluri10} or the structure of 
merger remnants \citep{HoffmanCDH10}. 

We have developed a new publicly available software tool, \SMILE%
\footnote{The acronym stands for ``Schwarzschild Modelling Interactive expLoratory Environment''. 
The software can be downloaded at \texttt{http://td.lpi.ru/\symbol{126}eugvas/smile/}.},
for orbit analysis and Schwarzschild modelling of triaxial stellar systems, 
intended to address a wide variety of ``generic'' questions from a theorist's perspective.
This paper presents an overview of various methods and algorithms (including some newly developed) 
used in the representation of potential, orbit classification, detection of chaos, and 
Schwarzschild modelling, that are implemented in \SMILE.

In the section~\ref{sec_SMILE_general} we introduce the basic definitions for the systems being 
modelled, and describe the main constituents and applications of the software. 
Section~\ref{sec_potential} presents the potentials that can be used in the modelling, including 
several flexible representations of an arbitrary density profile, 
section~\ref{sec_orbit_analysis} is devoted to methods for analysis of orbit class and its 
chaotic properties, and 
section~\ref{sec_schw_modelling} describes the Schwarzschild orbit superposition method itself.
In the remainder of the paper we explore the accuracy of our general-purpose potential expansions
(section~\ref{sec_accuracy_potentials}) and the efficiency of constructing a triaxial model 
using all variants of Schwarzschild method considered in the paper 
(section~\ref{sec_schw_modelling_tests}).

\section{The scope of the software}  \label{sec_SMILE_general}

\SMILE is designed in a modular way, allowing for different parts of the code to be used 
independently in other software (for instance, the potential represented as a basis-set expansion 
could be used as an additional smooth component in an external \Nbody simulation program 
\citep[e.g.][]{LowingJEF11}, 
or the orbit analysis could be applied to a trajectory extracted from an \Nbody simulation). 
There are several principal constituents, which will be described in more detail in 
the following sections. 

The first is the potential-density pair, which is used to numerically integrate the orbits 
and represent the mass distribution in Schwarzschild model. 
There are several standard analytical mass models and three more general representations of 
an arbitrary density profile, described in section~\ref{sec_potential}.
The second fundamental object is an orbit with given initial conditions, evolved for a given time 
in this potential. The orbit properties are determined by several orbit analysis and chaos 
detection methods, presented in section~\ref{sec_orbit_analysis}. 
The third constituent is the orbit library, which is a collection of orbits with the same energy, 
or covering all possible energies in the model.
In the first case, these orbits may be plotted on the Poincar\'e surface of section (for 2D 
potentials), or on the frequency map (for the 3D case, discussed in section~\ref{sec_freq_map}), 
to get a visual insight into the orbital structure of the potential. 
In the second case, this collection of orbits is the source component of the Schwarzschild model.
The target of the model may consist of several components: the density model corresponding to 
the potential in which the orbits were computed (there exist several possible representations 
of the density model, introduced in section~\ref{sec_schw_variants}), kinematic information 
(e.g. the velocity anisotropy as a function of radius), or the observational constraints in 
the form of surface density and velocity measurements with associated uncertainties. 
The latter case is not implemented in the present version of the software, 
but may be described within the same framework. 
The module for Schwarzschild modelling takes together the source and target components and 
finds the solution for orbit weights by solving an optimization problem 
(section~\ref{sec_schw_optimization}).

The tasks that can be performed using \SMILE include: 
\begin{itemize}
\item visualization and analysis of the properties of individual orbits;
\item exploration of the orbital structure of a given potential (one of well-studied analytical 
models, or an \Nbody snapshot with no \textit{a priori} known structure);
\item construction of self-consistent Schwarzschild models for the given density profile, with 
adjustable properties (e.g. velocity anisotropy or the fraction of chaotic orbits), and their 
conversion to an \Nbody representation.
\end{itemize}
Many of these tasks can be performed interactively (hence the title of the program) in 
the graphical interface; there is also a scriptable console version more suitable for large-scale 
computations on multi-core processors.

\section{Potential models}  \label{sec_potential}

A number of standard non-rotating triaxial mass models are implemented:
\begin{itemize}
\item Logarithmic: $\Phi(\tilde r) = \ln(R_c^2+\tilde r^2)$, 
which was studied in \citet{MiraldaSchw89,Schw93,PapLaskar98};
\item Triaxial generalization of \citet{Dehnen93} double power-law model: 
\begin{equation}  \label{eq_Dehnen_profile}
\rho(\tilde r) = \frac{3-\gamma}{4\pi\, p q} \tilde r^{-\gamma} (1+\tilde r)^{-(4-\gamma)} , 
\end{equation}
studied extensively by a number of authors 
\citep{MerrittFridman96, ValluriMerritt98, WachlinFerraz98, Siopis99};
\item Scale-free (single power-law): $\rho(\tilde r) = \tilde r^{-\gamma}$, studied in \citet{Terzic02};
\item Anisotropic harmonic oscillator: $\Phi(\tilde r) = \tilde r^2$, used in \citet{KandrupSideris02}.
\end{itemize}
Here $\tilde r=(x^2 + y^2/q^2 + z^2/p^2)^{1/2}$ is the elliptical radius, and axis ratios 
$q=y/x$, $p=z/x$ ($p\le q\le 1$) are defined for the potential (in the case of logarithmic 
and harmonic potentials) or for the density (in other cases).
$x$, $y$ and $z$ are the longest, intermediate and short axes, correspondingly.
A central supermassive black hole may be added to any potential (actually it is modelled as 
a Newtonian potential, since general relativistic effects are presumably not important for 
the global dynamics in the galaxy).

In addition to these models, there are several more flexible options.
One is a representation of a potential-density pair in terms of a finite number of basis functions 
with certain coefficients, and evaluation of forces and their derivatives as a sum over these 
functions, for which analytical expressions exist.
For not very flattened systems, an efficient choice is to write each member of the basis set as 
a product of a function depending on radius and a spherical harmonic: 
\begin{equation}  \label{eq_basis_set}
\Phi(r,\theta,\phi) = \sum_{n=0}^{\nmax} \sum_{l=0}^{\lmax} \sum_{m=-l}^{l} 
  A_{nlm}\,\Phi_{nl}(r)\, Y_l^m(\theta,\phi) \,.
\end{equation}

The idea to approximate a potential-density pair by a finite number of basis functions with given 
coefficients goes back to \citet{CluttonBrock73}, who used a set of functions based on the Plummer 
model, suitable to deal with cored density profiles. 
Later, \citet{HernquistOstriker92} introduced another class of basis functions, adapted for 
the cuspy \citet{Hernquist90} profile, to use in their self-consistent field (SCF) N-body 
method. \citet{Zhao96a} introduced a generalized $\alpha$-model including two previous cases; 
there exist also other bases sets for near-spherical \citep{AllenPP90, RahmatiJalali09} or 
flattened near-axisymmetric \citep[e.g.][]{BrownPapaloizou98} systems.
We use the basis set of \citet{Zhao96a}, which is general enough to accomodate both cuspy and 
cored density profiles, with a suitably chosen parameter $\alpha$.
\citep[For a discussion on the effect of different basis sets see][]
{CarpinteroWachlin06, KalapotharakosEV08}.
This formalism and associated formulae are presented in the Appendix \ref{sec_app_bse}, and 
tests for accuracy and recipes for choosing $\alpha$ are discussed in 
Section~\ref{sec_accuracy_potentials}.
A variation of this approach is to numerically construct a basis set whose lowest order function 
is specifically tuned for a particular mass distribution, and higher order terms are derived 
according to a certain procedure involving orthogonalization of the whole set 
\citep{Saha93, BrownPapaloizou98, Weinberg99}.
We propose another, conceptually simple approach described below.

Instead of requiring radial functions to form a orthogonal basis set, we may represent 
them as arbitrary smooth functions, namely, splines with some finite number of nodes:
\begin{equation}  \label{eq_spline_basis}
\Phi(r,\theta,\phi) = \sum_{l=0}^{\lmax} \sum_{m=-l}^{l} 
  \Phi_{lm}(r)\, Y_l^m(\theta,\phi) \,.
\end{equation}

The advantage over the basis set approach is its flexibility -- one may easily adapt it to any 
underlying potential model; the number and radii of nodal points may be chosen arbitrary 
(however it is easier to have them equal for all spherical harmonics); the evaluation of potential 
at a given point depends only on the coefficients at a few nearby nodes rather than on the whole 
basis set. Derivatives of the potential up to the second order are continuous and easily evaluated.
A similar approach was used for Schwarzschild modelling in \citet{ValluriME04} and \citet{Siopis09} 
for an axially symmetric potential.
In the context of \Nbody simulations, a related ``spherical-harmonic expansion'' method, pioneered 
by \citet{Aarseth67, vanAlbada77, White83}, computes coefficients at each particle's location, 
optionally introducing softening to cope with force divergence as two particles approach each other 
in radius. In the variant proposed by \citet{McGlynn84, Sellwood03}, coefficients of angular 
expansion are evaluated at a small set of radial grid points; radial dependence of forces 
is then linearly interpolated between grid nodes while the angular dependence is given by 
truncated spherical harmonic expansion. 
This potential solver is also used in the made-to-measure method of \citet{deLorenzi07}.

Both basis-set (BSE) and spline expansions may be constructed either for a given functional form 
of density profile (not requiring the corresponding potential to be known in a closed form), 
or from a finite number of points representing Monte-Carlo sampling of a density model. 
In the former case, one may use an analytic density model from a predefined set 
(e.g. Dehnen, Plummer, isochrone, etc.), or a flexible smooth parametrization of an arbitrary 
density profile by a Multi-Gaussian expansion \citep{EmsellemMB94, Cappellari02}.
In the case of initialization from an \Nbody snapshot, the spline coefficients are 
calculated by a method suggested in \citet{Merritt96}, similar to non-parametric density 
estimators of \citet{MerrittTremblay94}: 
first the angular expansion coefficients are evaluated at each particle's radius, then 
a smoothing spline is constructed which approximates the gross radial dependence of these 
coefficients while smoothing local fluctuations. More details on the spline expansion method 
are given in the Appendix \ref{sec_app_spline}, and the tests are presented in 
Section~\ref{sec_accuracy_potentials}.
 
Yet another option is to use directly the potential of \Nbody system of ``frozen'' (fixed in 
place) bodies. We use the potential solver based on the \cite{BarnesHut86} tree-code algorithm. 
It can represent almost any possible shape of the potential, but is much slower and noisier 
in approximating a given smooth density profile. It may use a variable softening length, 
the optimal choice for which is discussed in Section~\ref{sec_accuracy_treecode}.

\section{Orbit analysis}  \label{sec_orbit_analysis}

\subsection{Orbit integration}  \label{sec_orbit_integration}

Orbit integration is performed by the DOP853 algorithm \citep{DOP853}, which is a 
8th order Runge-Kutta scheme with adaptive step size. 
It is well suited for a subsequent Fourier analysis of trajectory because of its ability 
to produce ``dense output'', i.e. interpolated values of the function at arbitrary (in particular, 
equally spaced) moments of time. The energy of orbit is conserved to the relative accuracy 
typically better than $10^{-8}$ per 100 dynamical times for all smooth potentials except the 
Spline expansion, which is only twice continuously differentiable and hence demonstrates 
a lower (but still sufficient for most purposes) energy conservation accuracy 
($\sim 10^{-5}-10^{-6}$) in the high-order integrator.

For the frozen \Nbody potential we use a leap-frog integrator with adaptive timestep selection 
and optional timestep symmetrization \citep{HutMM95} which reduces secular energy drift. 
The reason for using a lower-order integrator is that the potential of the tree-code is 
discontinuous: when a trajectory crosses a point at which a nearby tree cell is opened (i.e. 
decomposed into sub-elements), which occurs when the distance to the cell is smaller than the 
cell size divided by the opening angle $\theta$, the potentials of an unresolved and resolved cell 
do not match. Therefore, the energy of an orbit is not well conserved during integration, 
no matter how small timesteps are. The error in potential approximation rapidly decreases with 
decreasing $\theta$, however, computational cost also increases quickly. 
Overall, for $\theta\simeq 0.5$ the accuracy of energy conservation is $\sim10^{-3}$; 
for more discussion see \citet{BarnesHut89}. 

For a given value of energy the period of long($x$)-axis orbit with the same energy is calculated, 
and it is used as a unit of time (hereafter $\tdyn$, dynamical time) 
and frequency in the following analysis. 
(This is different from most studies that use the period of closed loop orbit in $x-y$ plane 
for this purpose, but our definition has the advantage of being the longest possible period 
of any orbit with a given energy). 

\begin{figure} 
$$\includegraphics[width=6cm]{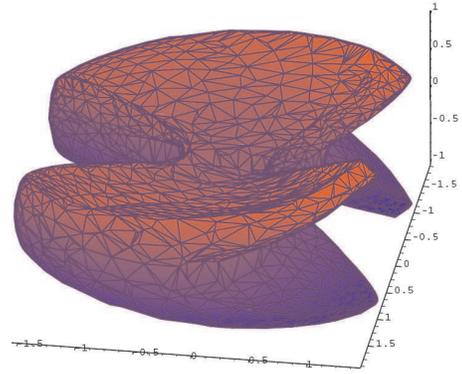} $$
\caption{ \small 3D rendering of a (2,1,-2) thin orbit.
} \label{fig_orbit}
\end{figure} 

Visualization of orbits is quite an important tool; orbit may be rendered either in projection 
on one of the three principal planes or in 3D. Additionally, an algorithm for rendering orbit 
as a solid body is implemented, which is based on Delaunay tesselation of set of points comprising 
the orbit, and removal of hidden surfaces to leave out only the outer boundary of the volume that 
the orbit fills (Fig.~\ref{fig_orbit}).

\subsection{Frequency analysis and orbit classification}  \label{sec_freq_analysis}

Orbits are classified by their spectra using the following scheme, 
based on \citet{CarpinteroAguilar98} (which, in turn, is an improvement of 
the method proposed by \citet{BinneySpergel82}). 
First we obtain the complex spectra of each spatial coordinate $x_c(t)$ by Fourier transform.
Then we extract the most prominent spectral lines $\omega_{c,j}$ for each coordinate: 
for each line, its amplitude and phase is determined using the Hunter's DFT method 
\citep{Hunter02}; other studies used similar techniques based on Laskar's NAFF algorithm 
\citep{Laskar93, ValluriMerritt98} or FMFT \citep{SidlNesv97}. 
All these methods employ Hanning window filtering on input data, which enhances the accuracy 
of frequency determination relative to the simple Fourier transform, used in the pioneering work 
of \citet{BinneySpergel82}. 
The contribution of each detected line to the complex spectrum is subtracted and the process is 
repeated until we find ten lines in each coordinate or the amplitude of a line drops below 
$10^{-2}$ times the amplitude of the first line.
Finally, the orbit classification consists in analyzing the relations between $\omega_{c,j}$.

A regular orbit in three dimensions should have no more than three fundamental frequencies 
$\Omega_k$, so that each spectral line may be expressed as a sum of harmonics of these fundamental 
frequencies with integer coefficients: $\omega_{c,j} = \sum_{k=1}^3 a_{cjk}\Omega_k$.
For some orbits the most prominent lines%
\footnote{Our choice of unit frequency makes it possible to put the LFCC in the correct order
$1\le \omega_x \le \omega_y \le \omega_z$, even if these lines are not the largest in amplitude.}
$\omega_{c,1} \equiv \omega_c$ in each coordinate $c$ 
(called LFCCs -- leading frequencies in Cartesian coordinates) coincide with the fundamental 
frequencies, these orbits are non-resonant boxes. 
For others, two or more LFCCs are in a resonant $m\!:\!n$ relation ($m\omega_{c1}=n\omega_{c2}$ 
with integer $m,n$). These orbits belong to the given resonant family, in which the parent orbit 
is closed in the $c_1-c_2$ plane. The additional fundamental frequency corresponds to the 
libration about this parent orbit, and is given by the difference between frequencies of 
the main and satellite lines. 
Besides this, in 3D systems there exist an important class of thin orbits with the three 
leading frequencies being linearly dependent (with integer coefficients):
$l\omega_1 + m\omega_2 + n\omega_3 = 0$.
These are labelled as $(l,m,n)$ thin orbits (of the three numbers at least one is negative) 
and are easily identified on a frequency map (see section~\ref{sec_freq_map}). 
In fact, resonant orbits are a subclass of thin orbits: $(m,-l,0)$ thin orbit may be 
alternatively termed as $l\!:\!m\!:*$ resonance, and $(mn,nl,-lm)$ orbit is named $l\!:\!m\!:\!n$ 
resonance.
The origin of term ``thin'' lies in the fact that a parent orbit of such an orbit family
is indeed confined to a 2D surface in the configuration space (possibly self-crossing).
This parent orbit has only two fundamental frequencies, and in the case of a closed orbit 
there is only one frequency. All orbits belonging to the associated family of a particular thin 
orbit have additional fundamental frequencies which may be viewed as libration frequencies around 
the parent orbit.
These thin orbit families are particularly important for triaxial models with cuspy density 
profiles, as they replace classical box orbits and provide the structure of the phase space 
\citep{MerrittValluri99}.

The most abondant subclass of resonant orbits are tube orbits, which have a 1:1 relation 
between frequencies in two coordinates. They have a fixed sense of rotation around the remaining 
axis. Accordingly, *:1:1 resonant orbits are called LAT (\hbox{$x$-,} or long-axis tubes), 
and 1:1:* orbits are SAT (\hbox{$z$-,} short-axis tubes); there are no stable tube orbits around 
the intermediate axis.
However, some of chaotic orbits may also have 1:1 correspondence between leading frequencies, 
but may not have a definite sense of rotation; in this case they are not labelled as tube orbits.

All other orbits which are not tubes, thin orbits or resonances, are called box orbits%
\footnote{In the vicinity of the central black hole, box orbits are replaced by regular 
pyramid orbits \citep{MerrittValluri99}.}.
Classification in two dimensions is similar but simpler: there exist only boxes and $m$:$n$ 
resonances (of which 1:1 are tubes), which also may be regular or chaotic. 

\begin{figure} 
$$\includegraphics{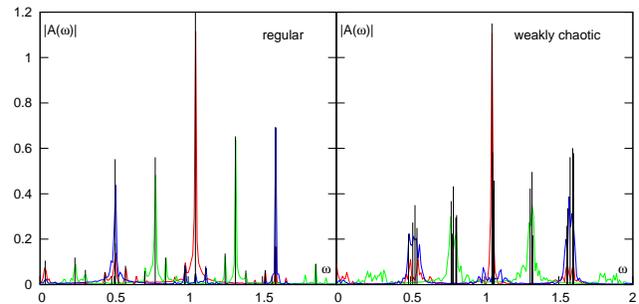} $$
\caption{ Spectral analysis of two orbits with slightly different initial conditions in 
$\gamma=1$ Dehnen model. 
The left one is $1:2:-1$ regular thin tube orbit, the right is weakly chaotic, sticky orbit 
in the vicinity of the same resonance. Shown in red/green/blue are amplitudes of complex spectra 
of motion in x/y/z coordinate, and vertical lines denote identified lines. 
Although the spectra look similar at first sight (when looking at amplitudes only), 
clustering of lines at the same frequency in the right panel hints at the chaotic nature 
of the orbit: after the subtraction of a line from the complex spectrum, its amplitude decreases 
only weakly, and the next line is found at almost the same frequency. By contrast, in the left 
panel all satellite lines are clearly identified and distinct, and the whole spectrum is a sum 
of linear combinations of just three fundamental frequencies with integer coefficients.
} \label{fig_spectrum}
\end{figure} 

Those orbits which display more complicated spectra (not all lines are expressed as linear 
combination of two/three frequencies) are additionally labelled chaotic (based on analysis of 
spectra), although this criterion for chaos is less strict than those discussed below.
Note that we do not have a special class for chaotic orbits: they fall into the most appropriate 
basic class (usually box, but some weakly chaotic orbits are also found among tubes and other 
resonances).

There is an additional criterion for chaoticity of an orbit: if we take the difference between 
corresponding fundamental frequencies $\Omega_c^{(1)}$ and $\Omega_c^{(2)}$ 
calculated on the first and second halves of the orbit, then this ``frequency diffusion rate'' 
(FDR) is large for chaotic orbits and small for regular ones \citep{Laskar93}. 
We use the LFCC diffusion rate, defined as the average relative change of three frequencies:
\begin{equation}  \label{eq_lfccdiff}
\Delta\omega = \frac{1}{3} \sum_{c=1}^3 
  \frac{|\omega_c^{(1)} - \omega_c^{(2)}|}{(\omega_c^{(1)} + \omega_c^{(2)})/2} \;.
\end{equation}

To account for the possibility of misinterpretation of two lines with similar amplitudes, 
frequencies that have relative difference greater than 0.5 are excluded from this averaging. 
Fig.~\ref{fig_spectrum} (right panel) shows an example of a weakly chaotic 
($\Delta\omega\sim 10^{-2}$) orbit in the vicinity of the $1\!:\!2\!:\!-1$ resonance, for which 
the leading frequencies do not form isolated peaks but rather clusters of lines, demonstrating 
that the complex spectrum is not described by just one line in the vicinity of a peak. 
Consequently, lines in these clusters change erratically in amplitude and position between 
the first and last halves of the orbit, which contributes to the rather high value of 
$\Delta\omega$. 

The FDR is, unfortunately, not a strict measure of chaos, nor is it well defined by itself.
The spectrum of values of FDR is typically continuous with no clear distinction between 
regular and chaotic orbits, which in part is due to the existence of ``sticky'' chaotic orbits
\citep[e.g.][]{ContopoulosHarsoula10}, which resemble regular ones for many periods, and 
consequently have low FDR. 
If one computes the FDR over a longer interval of time, these sticky orbits may become unstuck 
and demonstrate a higher value of FDR.
On the other hand, a perfectly regular orbit may sometimes have a rather large FDR because 
of two very close spectral lines with comparable amplitudes, which both degrades the accuracy 
of frequency determination and increases the time required for an orbit to fill its invariant 
torus uniformly. Over a longer interval, these nearby lines would be better resolved and 
such an orbit would attain a substantially lower FDR, which for regular orbits is ultimately 
limited by the accuracy of energy conservation.

\begin{figure} 
$$\includegraphics{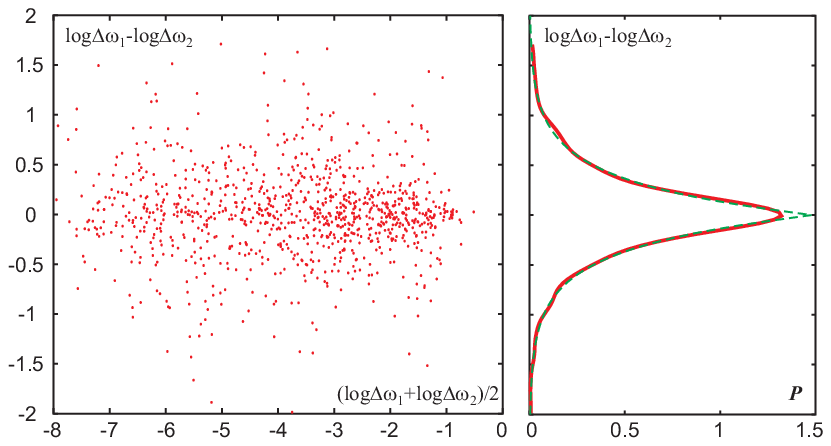} $$
\caption{ Correspondence between frequency diffusion rates $\Delta\omega_1$ and $\Delta\omega_2$, 
calculated for two intervals: [0:100] and [10:110] $\tdyn$, for a sample of $10^3$ orbits 
from a triaxial $\gamma=1$ Dehnen model. \protect\\
\textbf{Left panel:} difference $\log\Delta\omega_1-\log\Delta\omega_2$ plotted against 
the average value $(\log\Delta\omega_1+\log\Delta\omega_2)/2$. \protect\\
\textbf{Right panel:} probability distribution function of this difference (solid line); 
for comparison the Laplace distribution function $\exp(-|x|/\delta)/(2\delta)$ is shown 
(dashed line), for the value of dispersion $\delta=1/3$.
It demonstrates that $\Delta\omega$ is not a strictly defined quantity; it has variations 
of $\pm 0.3$ orders of magnitude.
} \label{fig_fdr_fuzziness}
\end{figure} 

A more fundamental problem is that even for the same time span the FDR is not a strictly defined 
quantity itself, i.e.\ it may vary by a factor of few when measured for two successive time 
intervals. This can be understood from the following simplified argument: 
suppose the ``instant'' value of frequency is a random quantity with a mean value $\omega_0$ and 
a dispersion $\delta \omega$, and the frequency measured over an interval is simply an average 
of this quantity. 
Then the FDR over any interval is a random quantity with dispersion 
$\sim \sqrt{2}\delta \omega$, or, if we take the distribution of $\log |\Delta \omega|$, 
it will be peaked around $\log \delta\omega$ with a scatter of $\sim 1/2$ dex. 
Similar uncertainty relates the values of FDR calculated for two different intervals of time. 
Indeed, for a particular case of a triaxial Dehnen potential we find that the correspondence 
between $\log \Delta \omega$ measured for two different intervals of the same orbit 
is well described by the following probability distribution: 
$P(\log\Delta \omega^{(1)}-\log\Delta \omega^{(2)} = X) = \frac{3}{2}\exp(-3|X|)$ 
(Fig.~\ref{fig_fdr_fuzziness}). 
To summarize, FDR is an approximate measure of chaos with uncertainty of $0.3-0.5$ orders of 
magnitude. Yet it correlates with the other chaos indicator, the Lyapunov exponent. 

\subsection{Lyapunov exponent}  \label{sec_lyapunov}

\begin{figure} 
$$\includegraphics[width=8cm]{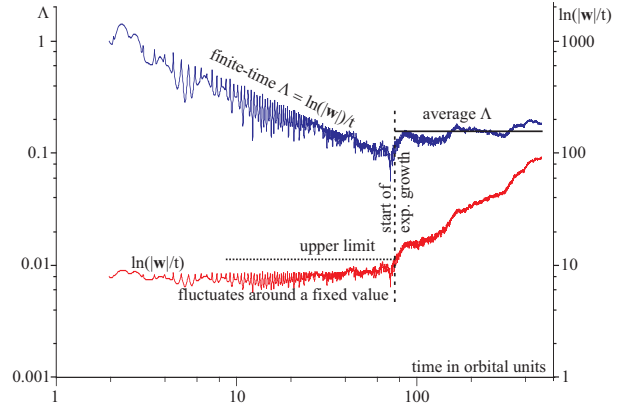} $$
\caption{ The scheme for estimating Lyapunov exponent $\Lambda$. 
In the period of linear growth of deviation $\boldsymbol{w}$, it fluctuates about a fixed value, 
and the estimate for $\Lambda$ falls as $t^{-1}$. 
When $\boldsymbol{w}$ starts to grow faster, the exponential regime is triggered, and $\Lambda$ 
is estimated as an average on the exp growth interval.
If no such growth is detected then $\Lambda$ is assumed to be zero.
} \label{fig_lyapunov}
\end{figure} 

The Lyapunov exponent (or, more precisely, the largest Lyapunov exponent) $\Lambda$ 
is another measure of chaoticity of an orbit. 
Given a trajectory $\boldsymbol{x}(t)$ and another infinitely close trajectory 
$\boldsymbol{x}+\boldsymbol{w}(t)$, we follow the evolution of deviation vector $\boldsymbol{w}$. 
For a regular orbit, the magnitude of this vector, averaged over some time interval longer than 
the orbit period, grows at most linearly with $t$; for a chaotic one it grows exponentially, and 
\begin{equation}  \label{eq_lambda}
\Lambda \equiv \lim_{t \to \infty} (\ln|\boldsymbol{w}|)/t \,.
\end{equation}

The usual method of computing $\Lambda$ is integration of the variational equation 
\cite[e.g.][]{Skokos10} along with the orbit, or simply integration of a nearby orbit. 
While the first method is more powerful in the sense that it may give not only the largest, 
but in principle the whole set of Lyapunov numbers \citep{UdryPfenniger88}, 
it requires the knowledge of the second derivatives of the potential%
\footnote{Eq.12 in \citet{MerrittFridman96} for the second derivative of triaxial Dehnen potential 
contains a typo, there should be $a_j^2-a_i^2$ in the denominator, instead of $+$.}.
The second method is relatively straightforward -- one needs to integrate the same equations of 
motions twice, and compute the deviation vector as the difference between the two orbits. 
The only issue is to keep $\boldsymbol{w}$ small, that is, the orbits must stay close despite the 
exponential divergence. Thus $\boldsymbol{w}$ should be renormalized to a very small value each 
time it grows above certain threshold (still small enough, but orders of magnitude larger than 
initial separation); however, one should be careful to avoid false positive values of $\Lambda$ 
that may appear due to roundoff errors \citep{Tancredi01}, in particular, for orbits that come 
very close to the central black hole.

In practice, one may calculate only the finite-time approximation for the true Lyapunov exponent 
for a given integration time. Eq.~\ref{eq_lambda} shows that such a finite-time estimate for 
a regular orbit decreases as $\sim t^{-1}$; therefore, the usually adopted approach is to 
find a threshold value for the finite-time estimate of $\Lambda$ that roughly separates regular 
and chaotic orbits. 
We use the following improved method that does not require to define such a threshold and gives 
either a nonzero estimate for $\Lambda$ in the case of a chaotic orbit, or zero if the orbit 
was not detected to be chaotic.
For a regular orbit -- or for some initial interval of a chaotic orbit -- $\boldsymbol{w}$ 
grows linearly, so that $|\boldsymbol{w}|/t$ fluctuates around some constant value 
(the period of fluctuations corresponds to characteristic orbital period; to eliminate these 
oscillations, we use a median value of $|\boldsymbol{w}|$ over the interval $2\tdyn$)%
\footnote{This method does not work well in some degenerate cases such as the harmonic potential, 
in which $\boldsymbol{w}$ on average does not grow at all.}.
When (and if) the exponential growth starts to dominate, $\Lambda$ may be estimated as the average 
value of $(\ln|\boldsymbol{w}|)/t$ over the period of exponential growth. 
If no such growth is detected, then $\Lambda$ is assumed to be zero (or, more precisely, an upper 
limit may be placed). 
The exponential growth regime is triggered when the current value of $|\boldsymbol{w}|/t$ is several 
times larger than the average over previous time. 
In addition, we normalize $\Lambda$ to the characteristic orbital frequency, so that its value 
is a relative measure of chaotic behaviour of an orbit independent of its period 
\citep[so-called specific finite-time Lyapunov characteristic number,][]{VoglisKS02}.
Our scheme is summarized in Fig.~\ref{fig_lyapunov}.
One should keep in mind that the ability of detecting chaotic orbits by their positive Lyapunov 
exponent depends on the interval of integration: for weakly chaotic orbits the exponential growth 
starts to manifest itself only after a long enough time. The finite-time estimate for $\Lambda$ 
may also depend on the integration time because the growth of $\boldsymbol{w}$ is not exactly 
exponential and exhibits fluctuations before reaching asymptotic regime.


\begin{figure} 
$$\includegraphics{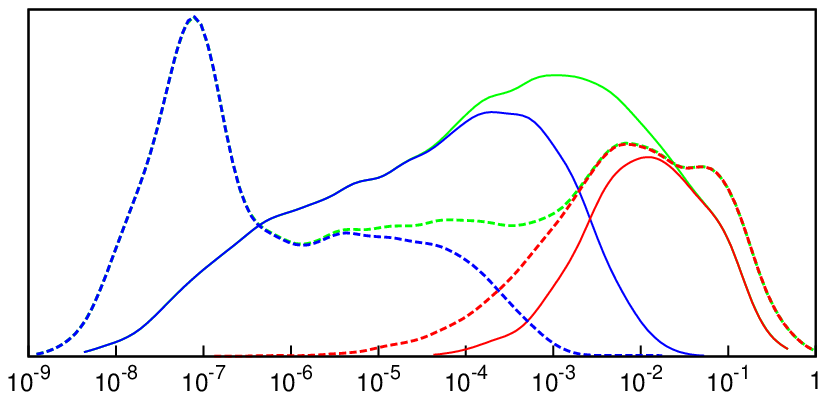} $$
$$\includegraphics[angle=90]{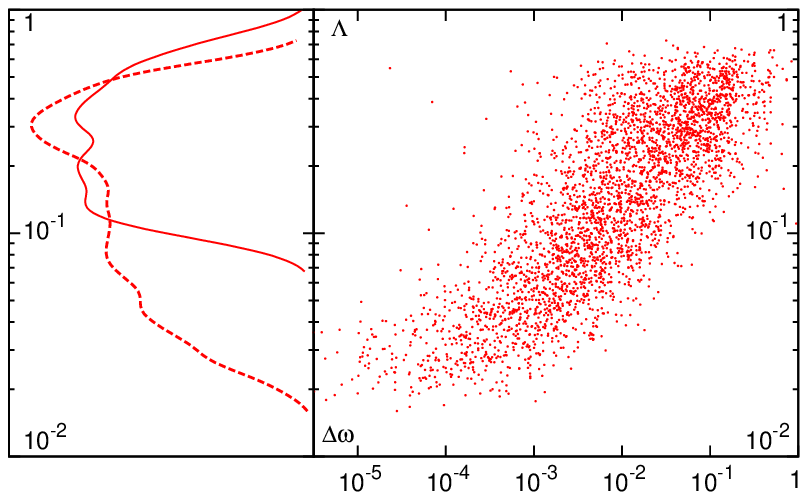} $$
\caption{ 
Correlation between chaotic properties of orbits in a triaxial $\gamma=1$ Dehnen model. 
Shown is an ensemble of $10^4$ orbits, sampled randomly from all energies and integrated for 
100 (solid lines) and 500 (dashed lines) $\tdyn$. \protect\\
\textbf{Top}: histograms of frequency diffusion rates (FDR) $\Delta\omega$ for orbits that appear 
to be regular (blue, left) or chaotic (red, right), based on their Lyapunov exponents $\Lambda$ 
being zero or non-zero. 
The rather clear separation in $\Delta\omega$ (limited by the uncertainty of FDR determination, 
see Fig.~\ref{fig_fdr_fuzziness}) between the two sorts of orbits suggests an approximate 
threshold $\Delta\omega_\mathrm{ch}$ for chaos detection based on FDR;
for the orbit ensemble shown here, $\Delta\omega_\mathrm{ch} \sim 10^{-3}$ for 100 
$\tdyn$ and $10^{-4}$ for 500 $\tdyn$. \protect\\
\textbf{Bottom left}: histogram of Lyapunov exponents (only orbits with nonzero $\Lambda$ are 
shown, which comprise 30\% (43\%) of all orbits for 100 (500) $\tdyn$). \protect\\
\textbf{Bottom right}: crossplot of $\Lambda$ and $\Delta\omega$ for orbits integrated for 
500 $\tdyn$. (Orbits with zero Lyapunov exponent are not shown). 
A correlation between the two chaos indicators is apparent.
} \label{fig_fdr_lambda_compare}
\end{figure} 

The two chaos indicators -- the frequency diffusion rate (FDR) $\Delta \omega$ (\ref{eq_lfccdiff}) 
and the Lyapunov exponent $\Lambda$ (\ref{eq_lambda}) -- are based on different methods yet 
demonstrate a rather good agreement in the chaos detection 
(see e.g.\ \citet{MaffioneDCG13} for a detailed comparison of various chaos indicators based on 
variational equation and on frequency analysis).
Fig.~\ref{fig_fdr_lambda_compare} presents a comparisons between $\Delta\omega$ and $\Lambda$ 
for $10^4$ orbits in a particular triaxial $\gamma=1$ Dehnen model.
It demonstrates that orbits labelled as regular or chaotic, based on $\Lambda$, have quite well 
separated distributions in $\Delta\omega$, with the overlap being comparable to the intrinsic 
uncertainty of FDR determination. Moreover, for orbits with $\Lambda>0$ there is a clear 
correlation between the two chaos indicators.
Meanwhile, the distribution in $\Delta\omega$ and $\Lambda$ is quite different for intervals 
of 100 and 500 $\tdyn$, as explained above, and the threshold $\Delta\omega_\mathrm{ch}$ 
separating regular and chaotic orbits does depend on the integration time.

\subsection{Frequency map}  \label{sec_freq_map}

\begin{figure} 
$$\includegraphics[angle=-90,width=8cm]{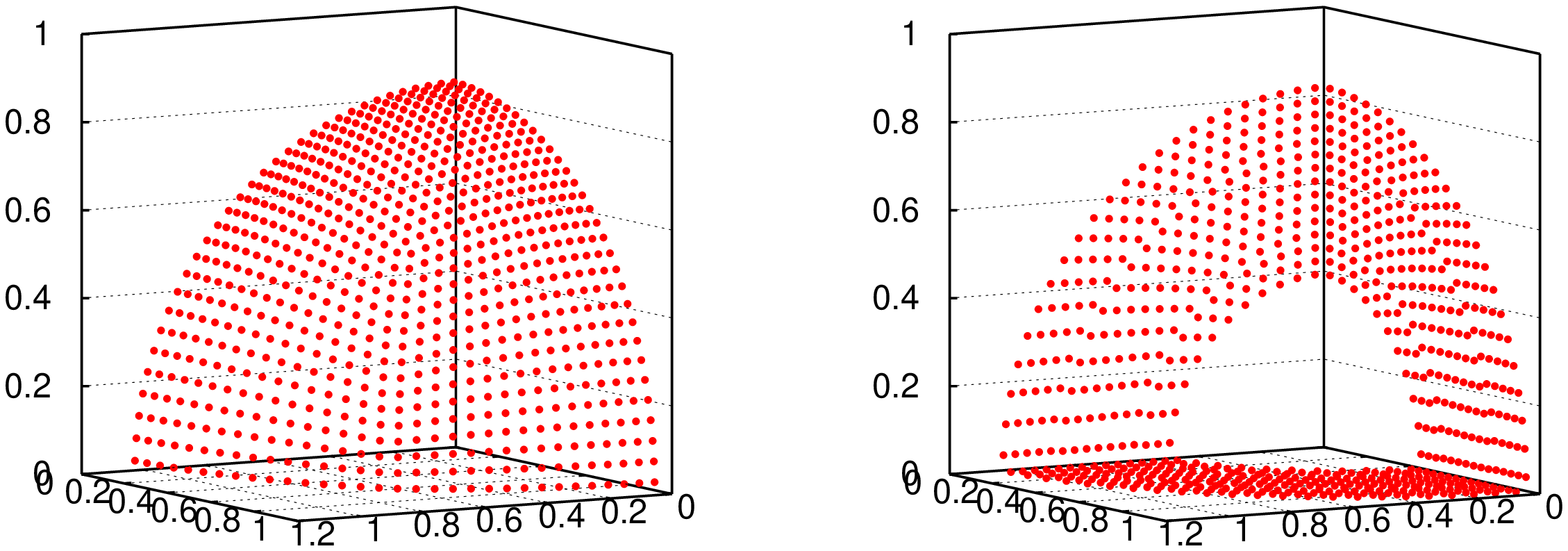} $$
$$\includegraphics{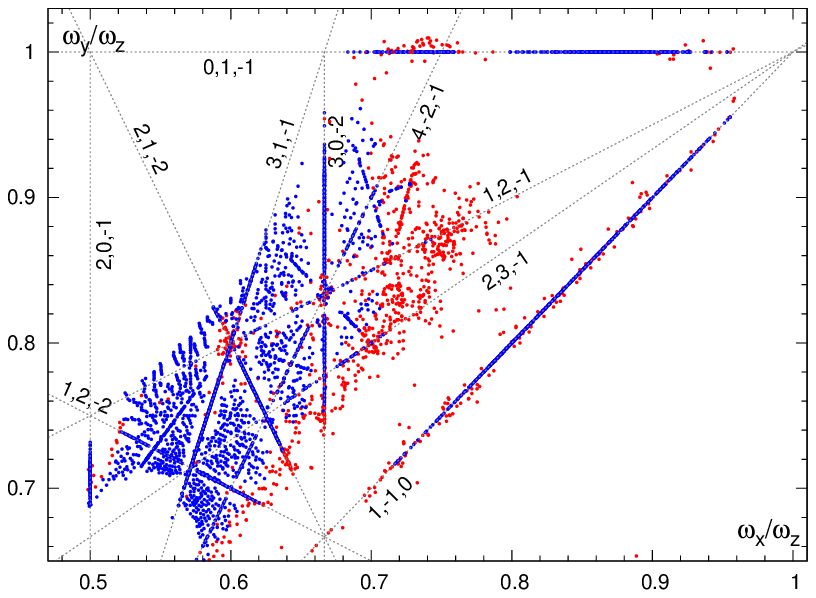} $$
\caption{ 
\textbf{Top:} Start-spaces for orbit library: left -- stationary (on the equipotential surface), 
right -- principal-plane (on $x-y$, $y-z$ and $z-x$ planes outside the periodic 1:1 orbit). 
\protect\\
\textbf{Bottom:} Frequency map for a $\gamma=0.75$ triaxial Dehnen models with $q=0.8,p=0.5$ at 
the energy $E=-0.5$.
Blue dots mark regular and red dots -- chaotic orbits (as determined by the value of Lyapunov 
exponent). Numerous resonant orbit families are clearly visible as lines, and regular non-resonant 
orbits form a quasi-regular pattern according to their initial conditions.
} \label{fig_freq_map}
\end{figure} 

Frequency map is a convenient and illustrative tool for analysing orbital structure of a potential 
\citep{PapLaskar98, ValluriMerritt98, WachlinFerraz98}. 
For 3D system we plot the ratio of LFCCs: $\omega_x/\omega_z$ vs. $\omega_y/\omega_z$
for a set of orbits, usually with regularly defined initial conditions. 
The points corresponding to resonant or thin orbits then group along certain lines on the map. 
Since they are very important in the dynamical structure of the potential,
this fact alone serves as an illustration of the orbital structure. 

Usually the map is constructed for a set of orbits of fixed energy, in which initial conditions 
for orbits are drawn from some start space. 
There exist two widely used start spaces \citep{Schw93}: one is the stationary, which contains 
orbits that have at least one zero-velocity point (then by definition they touch equipotential 
surface), the other is the principal-plane, consisting of orbits which traverse one of the 
principal planes ($x=0$, $y=0$ and $z=0$) with velocity perpendicular to it. 
The equipotential surface and each of the three principal planes are sampled in a regular manner 
(Fig.~\ref{fig_freq_map}, top). 
A set of non-chaotic orbits whose initial conditions lie on a regular grid of points in 
the start space will then appear as a visibly regular structure on the frequency map. 
Chaotic orbits do not have well-defined intrinsic frequencies, hence they will randomly fill 
the map and contaminate the regular structure, so they are plotted in a different color 
(Fig.~\ref{fig_freq_map}, bottom). 
The frequency map helps to identify the regions of phase space which contain mostly regular 
or chaotic orbits and highlights the most prominent the resonances. 

The Poincar\'e surface of section \citep[e.g.][]{LL92} is another important tool for analysing 
orbital structure of a two-dimensional potential, as well as for studying orbits confined to 
one of the principal planes in a 3D potential. This tool is also implemented in \SMILE.

\section{Schwarzschild modelling}  \label{sec_schw_modelling}

\subsection{Spherical mass models}  \label{sec_schw_spherical}

Before discussing the Schwarzschild method itself, we first outline the formalism used to deal 
with spherical isotropic mass models with an arbitrary density profile. 
These models are constructed from an array of $\{r, M(r)\}$ pairs specifying the dependence of 
enclosed mass on radius. The mass, potential and other dynamical properties are represented as 
spline functions in radius (with logarithmic scaling and careful extrapolation to small and large 
radii beyond the definition region, in the way similar to the Spline potential approximation).
The unique isotropic distribution function is given by the Eddington inversion formula \citep{BT}:
\begin{equation}  \label{eq_Eddington}
f(E) = \frac{1}{\sqrt{8}\,\pi^2} \int_E^0 \frac{d^2\rho}{d\Phi^2} \frac{d\Phi}{\sqrt{\Phi-E}} \;.
\end{equation}

These spherical models are used throughout \SMILE in various contexts, in particular, to generate 
the initial conditions for the orbit library used to create Schwarzschild models: for the given 
triaxial potential, a spherically-symmetric approximation model is created and the initial 
conditions are drawn from its distribution function. Of course, one may also use such a model 
to create a spherically-symmetric isotropic \Nbody model with a given density profile 
(there is a separate tool, \texttt{mkspherical}, that does just that).
The approach based on the distribution function gives better results than using Jeans equations 
with a locally Maxwellian approximation to the velocity distribution function 
\citep{KazantzidisMM04}.
Of course, even the simplest variants of Jeans models can account for varying degree of velocity 
anisotropy, but there also exist methods to derive anisotropic distribution function for spherical 
models \citep[e.g.][]{Ossipkov79, Merritt85, Cuddeford91, BaesVanHese07}. 
As we use spherical models only as a ``seed'' to construct a more general Schwarzschild model, 
a simple isotropic distribution function is enough for our purposes.

Another application of these models is to study dynamical properties of an existing \Nbody 
snapshot, for instance, the dependence of dynamical and relaxation time on radius or energy. 
To create such a model from an \Nbody snapshot, we again use the penalized spline smoothing 
approach. That is, we first sort particles in radius and define a radial grid of roughly 
logarithmically spaced $N_\mathrm{grid}\sim 10-50$ points so that each interval contains 
at least $10-50$ of particles. 
Then the $M(r)$ profile is obtained by fitting a penalized spline to the scaled variable 
$\mu (\xi\equiv \log r) \equiv \log[M(r)/(M_\infty-M(r))]$. The degree of smoothing may be 
adjusted to obtain a distribution function that is not very much fluctuating (or at least doesn't 
become negative occasionally). 
An alternative approach would be, for instance, fitting a local power-law density profile whose 
parameters depend on radius and are computed from a maximum likelihood estimate taking into 
account nearby radial points \citep[e.g.][]{ChurazovTFGDVJBG09}.
The creation of spherical models from \Nbody snapshots is also implemented as a separate tool; 
in \SMILE the same is done using the intermediate step of basis-set or Spline potential 
initialization.

\subsection{Variants of Schwarzschild method}  \label{sec_schw_variants}

Each orbit in a given potential is a solution of CBE (\ref{eq_CBE}): the distribution function 
is constant over the region of phase space occupied by the orbit, provided that it was integrated 
for a sufficiently long time to sample this region uniformly. 
The essence of the Schwarzschild's orbit superposition method is to obtain the self-consistent 
solution of both CBE and the Poisson equation (\ref{eq_poisson}) by combining these individual 
elements with certain weights to reproduce the density profile consistent with the potential 
used to integrate the orbits, possibly with some additional constraints (e.g. kinematical data).
It is clear that such a superposition is sought only in the configuration space, i.e.\ involves 
only the density or the potential created by individual orbits; for instance, one may write 
the equation for the density 
\begin{equation}  \label{eq_superposition}
\rho(\boldsymbol{r}) = \sum_{o=1}^{N_o} w_o\, \rho_o(\boldsymbol{r}) \;,
\end{equation}
where $w_o \ge 0$ are the orbit weights to be determined, and each orbit has a density profile 
$\rho_o(\boldsymbol{r})$. This equation can be approximately solved by discretizing both the 
original density profile and that of each individual orbit into a sum of certain basis functions, 
thus converting the continuous equation (\ref{eq_superposition}) into a finite linear system.
The classical Schwarzschild method consists of splitting the configuration space into $N_c$ 
cells, computing the required mass $m_c$ in each cell from the original density profile, 
and recording the fraction of time $t_{oc}$ that $o$-th orbit spends in $c$-th cell. 
Then the linear system of equation reads
\begin{equation}  \label{eq_linear_system}
m_c = \sum_{o=1}^{N_o} w_o\,t_{oc} \;, c=1..N_c \,.
\end{equation}

We may generalize the above definition to replace cells with some arbitrary constraints 
to be satisfied exactly or as closely as possible. Two such alternative formulations 
naturally arise from the definitions of basis-set (BSE) and Spline potential expansions.
Namely, we may use the coefficients of potential expansion of the original model as target 
constraints $m_c$, compute the expansion coefficients from the mass distribution of each orbit 
as $t_{oc}$ and then find the orbit weights $w_o$ so that the weighted sum of these coefficients 
reproduces the total potential. 
The linear character of orbit superposition is preserved in the potential expansion formalism. 
Accordingly, we introduce two additional variants of Schwarzschild models, named after 
the potential expansions: the BSE model and the Spline model; the original, grid-based 
formulation is called the Classic model.
The BSE model is analogous to the self-consistent field method of \citet{HernquistOstriker92}, 
with the difference that the coefficients are built up by summing not over individual particles, 
but over entire orbits.

An important difference between the Classic and the two new variants of the Schwarzschild 
method is that in the latter case, the basis functions of density expansion and the elements 
of the matrix $t_{oc}$ and the vector $m_c$ in (\ref{eq_linear_system}) are not necessarily 
non-negative. In other words, a single orbit with rather sharp edges, when represented by a 
small number of expansion coefficients, does not necessarily have a positive density everywhere; 
however, when adding up contributions from many orbits the resulting density is typically 
well-behaved (as long as the target expansion had a positive density everywhere).
On the other hand, the classical Schwarzschild method ensures only that the \textsl{average} 
density within each cell is equal to the required value, and does not address the issue of 
continuous variation of density across grid boundaries. The basis functions in the 
classical formulation are $\sqcap$-shaped functions with finite support and sharp edges, 
while in the proposed new variants these are smooth functions.
Recently \citet{JalaliTremaine11} suggested another generalization of the Schwarzschild method 
(although presently only for the spherical geometry), 
in which density, velocity dispersion and other quantities are represented as expansions over 
smooth functions with finite support, and additional Jeans equation constraints are used to 
improve the quality of solution. 

\begin{figure} 
$$\includegraphics[angle=-90,width=4.5cm]{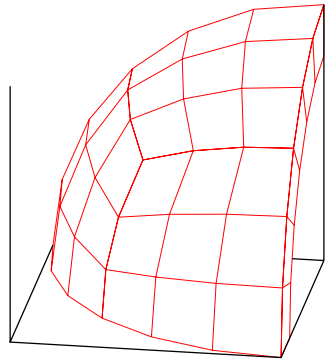} $$
\caption{ 
The configuration space grid in the Classic Schwarzschild model. Shown is one shell divided 
into three sectors, each sector is further divided into $n_\mathrm{seg}^2$ curvilinear rectangles 
(9 in this example).
} \label{fig_schw_grid}
\end{figure} 

The partitioning of configuration space in the Classic model is done in the similar way as 
in \citet{Siopis99}. We define $n_\mathrm{shell}+1$ concentric spherical shells at radii $r_s$ 
(the last shell's outer boundary goes to infinity, and this shell is not used in modelling). 
By default, shells are spaced in radii to contain approximately the same mass each, but 
this requirement is not necessary, and one may build a grid with a refinement near the center. 
Shells are further divided by the following angular grid. The sphere is split into three sectors 
by planes $x_i=x_j$; then each sector is divided into $n_\mathrm{seg} \times n_\mathrm{seg}$ 
rectangles by lines $x_i/x_j = \tan(\frac{\pi}4 n/n_\mathrm{seg})$ ($n=1..n_\mathrm{seg}$). 
This way we get $3\, n_\mathrm{seg}^2\, n_\mathrm{shell}$ cells (Fig.~\ref{fig_schw_grid}). 
The time $t_{oc}$ that an orbit spends in each cell is calculated with great precision thanks to 
the dense output feature: if two subsequent points on a trajectory fall into different cells, 
we find the exact moment of cell boundary crossing by nested binary divisions of the 
interval of time (this may in turn reveal a third cell in between, etc.).
This approach is more straightforward and precise than used in \cite{Siopis99, Terzic02}.

In the BSE and Spline models, we simply compute coefficients of expansion for each orbit, 
so that there are $n_\mathrm{rad} \times (l_\mathrm{max}/2+1) \times (l_\mathrm{max}/2+2)/2$ 
constraints in the model, where $n_\mathrm{rad}$ is the number of radial basis functions 
in the BSE model or the number of radial points in the Spline models (which are chosen 
in the same way as the concentric shells in the Classic model), and $l_\mathrm{max}$ is 
the order of angular expansion (the factor 1/2 comes from using only the even harmonics).
Unlike the initialization of the Spline potential from a set of discrete points, we do not 
perform penalized spline smoothing for orbit coefficients, to keep the problem linear.

In all three variants we also constrain the total mass of the model, and may have 
additional (e.g. kinematic) constraints; at present, one fairly simple variant of 
kinematic data modelling is implemented, which sets the velocity anisotropy profile 
as a function of radius. 
The configuration space is split into shells and the mean squared radial and transversal 
velocity ${v_r^2}_{,os}$ and ${v_t^2}_{,os}$ of each orbit in each shell is recorded. 
Then the following quantity is used as a constraint required to be zero:
\begin{equation}
\sum_{o=1}^{N_o} w_o\,({v_t^2}_{,os} - 2(1-\beta_s){v_r^2}_{,os}) \,.
\end{equation}

Here $\beta_s$ is the required value of the velocity anisotropy coefficient 
$\beta \equiv 1-\sigma_t^2/2\sigma_r^2$ \citep{BT} in the given radial shell.
In the practical application of the Schwarzschild method to the modelling of individual 
galaxies, the projected velocity distribution function is usually constrained; 
this is easy to add in the general formalism implemented in \SMILE.

Traditional approach to the construction of the orbit library is the following: 
choose the number of energy levels (typically corresponding to potential at the intersection 
of $N$-th radial shell with $x$ axis) and assign initial conditions at each energy shell 
similarly to that of frequency map (stationary and principal-plane start-spaces). 
The outer shell may be taken as either equipotential or equidensity surface, whichever 
is rounder, so that all finite spatial cells are assured to be threaded by some orbits.
However, we found that such discrete distribution in energy and in initial positions 
is not welcome when translating the Schwarzschild model to its \Nbody representation. 
Step-like distribution in energy tends to relax to continuous one, which introduces 
systematic evolution \citep{VasilievAthanassoula12}; 
in addition, the coarse radial resolution ($n_\mathrm{shell} \sim 20-50$) does not allow 
to sample the innermost particles well enough. 
Therefore, we draw the initial conditions for the orbit library from the spherical 
mass model discussed in the previous section, constructed for the given potential.

\subsection{Solving the optimization problem}  \label{sec_schw_optimization}

The linear system (\ref{eq_linear_system}) may be reformulated as an optimization problem, 
introducing auxiliary variables $\delta_c$ which are deviations between the required and 
the actual constraint values, normalized to some scaling constants $\eta_c$:
\begin{equation}  \label{eq_optim_delta}
\delta_c = \left(\sum_{o=1}^{N_o} w_o\,t_{oc} - m_c\right)/\eta_c \;, c=1..N_c \,.
\end{equation}

Obviously, we seek to minimize $|\delta_c|$, ideally making them zero, but in addition one may 
require that some other relations be satisfied as closely as possible, or to within a predefined 
tolerance, or that a certain functional of orbit weights be minimized 
(e.g. the sum of weights of chaotic orbits). 
A rather general formulation of this problem is to introduce an objective (penalty) function 
$\cal F$ and find $\mathrm{min}\,{\cal F}(\{w_o\})$, subject to the constraints $w_o \ge 0$. 
Various studies have adopted different forms for the penalty function and different methods to 
find the minimum.

One could incorporate the requirement of exact match between required and calculated mass in each 
cell (set $\delta_c=0$ as linear constraints); however, this is not physically justified -- given 
a number of other approximations used in modelling, sampling orbits etc., it is unreasonable to 
require strict equality. Instead, one may penalize the deviation of $\delta_c$ from zero and 
search for the solution that minimizes this penalty. (The scaling coefficients $\eta_c$ may be 
taken as some ``typical'' values of $m_c$, to give roughly similar significance to the deviations
$\delta_c$). This may be done in various ways: for instance, one may take 
\begin{equation}  \label{eq_penalty_quadratic}
{\cal F} = \alpha\sum_{c=1}^{N_c} \delta_c^2 + {\cal F}_\mathrm{additional} \,,
\end{equation}
(this is called non-negative least-square method, used in \citet{MerrittFridman96, Zhao96b}), 
or introduce additional non-negative variables and use
\begin{equation}  \label{eq_penalty_linear}
{\cal F} = \alpha\sum_{c=1}^{N_c} (\mu_c + \nu_c) + {\cal F}_\mathrm{additional} \,;\;
\mu_c, \nu_c \ge 0,\; \delta_c=\mu_c-\nu_c ,
\end{equation}
this effectively reduces to ${\cal F}=\alpha\sum |\delta_c|$; the latter approach was used in 
\citet{Siopis99, Terzic02}. Here $\alpha$ is the penalty coefficient discussed below.
In principle, the two formulations do the same job -- if possible, reach the exact solution, 
if not, attain the ``nearest possible'' one. 
Another option is to drop the condition that $\delta_c=0$ and instead require that 
$|\delta_c| \le \beta |m_c|$, where $\beta$ is the allowed fractional deviation from 
the exact constraint value (for example, 1\%); this approach was adopted in \citet{vdBosch08}.
It may be reformulated as the standard non-negative optimization problem by introducing 
additional variables $\mu_c,\nu_c \ge 0$ and doubling the number of equations:
\begin{equation}  \label{eq_penalty_tolerance_range}
\delta_c=\mu_c-\nu_c\,,\; \mu_c+\nu_c=\beta |m_c| \,,\;c=1..N_c \,.
\end{equation}

We have implemented the last two approaches -- either a tolerance range defined by fractional 
constraint deviation $\beta$, or a linear term in the penalty function proportional to $\alpha$.
(In principle, a combination of both variants is trivial to implement).
The first approach is also easily formulated in terms of the quadratic optimization problem, 
but it involves a dense matrix of quadratic coefficients with size $N_o\times N_o$, and hence
is less practical from the computational point of view. However, such quadratic problems with 
a specific form of the penalty function are efficiently solved with the non-negative least 
squares method \citep{NNLS}.

When Schwarzschild modelling is used to construct representations of observed galaxies, 
one usually computes the observable quantities (surface density and line-of-sight velocity 
distribution as functions of projected position) in the model ($Q_{i,\mathrm{mod}}$) 
and minimize their deviation from the actual observations $Q_{i,\mathrm{obs}}$, 
normalized by the measurement unsertainties $\Delta Q_{i,\mathrm{obs}}$:
\begin{equation}  \label{eq_penalty_chi2}
\chi^2 = \sum_{i=1}^{N_i} 
  \left(\frac{Q_{i,\mathrm{mod}}-Q_{i,\mathrm{obs}}}{\Delta Q_{i,\mathrm{obs}}}\right)^2 
  + {\cal F}_\mathrm{additional} \,.
\end{equation}

Then one seeks to minimize $\chi^2$ and derive the confidence intervals of the model parameters 
based on standard statistical $\Delta \chi^2$ criteria. In this approach, the self-consistency 
constraints for the density model (\ref{eq_linear_system}) may be either included in the same way 
as the observational constraints, with some artificially assigned uncertainty $\Delta Q_i$ 
\citep[e.g.][]{ValluriME04}, 
or as tolerance intervals via (\ref{eq_penalty_tolerance_range}) \citep[e.g.][]{vdBosch08}.

An important feature of Schwarzschild modelling is that a solution to the optimization problem, 
if exists, is typically highly non-unique. In principle, the number of orbits that are assigned 
non-zero weights may be as low as the number of constraints (which is typically $\gtrsim 10$ 
times smaller than the total number of orbits $N_o$; however, in some studies the opposite 
inequality is true, e.g. \citet{Verolme02} had four times more constraints than orbits). 
While this is a solution to the problem in the mathematical sense, it is often unacceptable 
from the physical point of view: large fluctuations in weights of nearby orbits in phase space 
are almost always unwelcome. 
For this reason, many studies employ additional means of ``regularization'' of orbit 
weight distribution, effectively adding some functional of orbit weights 
${\cal F}_\mathrm{additional}$ to the penalty function $\cal F$ or $\chi^2$.
There are two conceptually distinct methods of regularization (which are not mutually
exclusive): 
``local'' try to achieve smoothness by penalizing large variations in weight 
for orbits which are close in the phase space, according to some metric; 
``global'' intend to minimize deviations of orbit weights from some pre-determined 
prior values (most commonly, uniform, or flat priors).

The first approach \citep[e.g.][]{Zhao96b, CrettonZMR99, Verolme02, vdBosch08},
minimizes the second derivatives of orbit weight as a function of initial conditions 
assigned on a regular grid in the start-space, 
in which the proximity of orbits is determined by indices of the grid nodes. 
Since in our code we do not use regularly spaced initial conditions, this method 
is not applicable in our case.
The second kind of regularization assumes some prior values $\tilde w_o$ for orbit 
weights (most commonly, uniform values $\tilde w_o = M_\mathrm{total}/N_o$, 
although \citet{MorgantiGerhard12} argue for on-the-fly adjustment of 
the weight priors in the context of made-to-measure modelling).
Then a penalty term is added to the cost function which minimizes the deviations 
of the actual weights from these priors. 
This could be done in various ways.
\citet{RichstoneTremaine88} introduced the maximum entropy method, in which 
${\cal F}_\mathrm{additional} = -\lambda S$, where $\lambda$ is the regularization 
parameter, and $S$ is the (normalized) entropy, defined as
\begin{equation}  \label{eq_entropy}
S \equiv -\frac{1}{M_\mathrm{total}} \sum_{o=1}^{N_o} w_o \ln(w_o/\tilde w_o) \;,
  \quad M_\mathrm{total} \equiv \sum_{o=1}^{N_o} w_o .
\end{equation}

This method was used in \citet{Gebhardt00, Thomas04, Siopis09}%
\footnote{Eq.44 in \cite{Thomas04} has a sign error in the definition of entropy.}.
Another possibility is to use a quadratic regularization term in the penalty function:
\begin{equation}  \label{eq_regularization_quadratic}
{\cal F}_\mathrm{additional} = \frac{\lambda}{N_o} \sum_{o=1}^{N_o} (w_o/\tilde w_o)^2
\end{equation}

This approach, used in \citet{MerrittFridman96, Siopis99, ValluriME04}, gives 
very similar results to the maximum entropy method, but is simpler from computational 
point of view, since the regularization term is quadratic in $w_o$ and not a 
nonlinear function as in (\ref{eq_entropy}).
We adopted this second variant with uniform weight priors $\tilde w_o$, 
which is not a bad assumption given that our method of assigning initial conditions 
populates the phase space according to the isotropic distribution function for 
a given density profile.

In the context of Schwarzschild modelling of observational data, the regularization 
is considered a necessary ingredient because the number of observational constraints 
is typically much smaller than the number of free parameters (orbit weights), 
so that some smoothing is desirable to prevent overfitting (fitting the noise 
instead of actual physical properties of the system). 
The amount of smoothing is then controlled by a regularization parameter 
$\lambda$ which scales the contribution of the penalty term to the cost function,
and there are standard statistical methods of determining the optimal value of 
this parameter \citep[e.g.\ cross-validation technique,][]{Wahba90}.
Most commonly, the regularization coefficient is chosen as to achieve 
maximal smoothing which still does not deteriorate the quality of the fit more 
than by some acceptable value of $\Delta\chi^2$.
In our application of Schwarzschild method to the construction of models with 
pre-defined, noise-free properties, the necessity of smoothing is not obvious 
a priori. The linear system (\ref{eq_linear_system}) either has no solutions 
or infinitely many solutions, and unless some nonlinear objective function is 
used, the solution will tend to have only $\sim N_c$ orbits with nonzero weights.
\citet{VasilievAthanassoula12} have shown that a solution with a larger number 
of orbits orbits (lower average orbit weight) is more stable in the \Nbody simulation 
(in addition to be smoother and more aesthetically pleasant), 
therefore it is preferable to use regularization to select a solution with 
$\mathcal{O}(N_o)$ rather than $\mathcal{O}(N_c)$ effective orbits from all 
possible set of solutions that satisfy all constraints.
Since we do not have any trade-off between fit quality and smoothness, 
the value of regularization parameter $\lambda$ itself is not important, 
only insofar as it should not outweight the penalties for constraint violation 
in (\ref{eq_penalty_linear}), parametrized by the coefficient $\alpha$.

An important issue in modelling is whether (and how) to include chaotic orbits into 
the orbit library. 
The generalized Jeans' theorem \citep[e.g.][]{Kandrup98} states that to satisfy CBE, 
the distribution function must be constant in every ``well-connected'' region of phase space,
whether it is an invariant torus defined by three integrals of motion for a regular orbit 
or a hypersurface of higher dimensionality for a stochastic orbit.
The difficulty is that in many cases the distinction between regular and chaotic orbits 
is very blurry, and some weakly chaotic orbits retain a quasi-regular character for many 
periods -- much longer than any timescale of the model -- before jumping into another part 
of their reachable region of phase space, thereby violating the assumption of time-invariance 
of each building block of the model.
To combat this, \citet{Pfenniger84} adaptively increased integration time for those orbits 
(presumably chaotic ones) that exhibited substantial variation of time spent in each cell 
during the integration; however, for practical purposes there should be an upper limit 
to this time, and it may not guarantee an adequate phase space coverage of sticky orbits. 
Alternatively, \citet{vdBosch08} proposed to use ``dithered'' orbits, starting a bunch of 
orbits with slightly different initial conditions and combining their density into one block 
to be used in the optimization routine, which increases the coverage of phase space available 
for a given (slightly perturbed) orbit. 
We have experimented with this approach but did not find it to be superior to just 
using equivalently larger number of separate orbits in the solution.
Another interesting option is suggested by \citet{SiopisKandrup00}, who found that adding 
a weak noise term to the equations of motion substantially increases the rate of chaotic 
diffusion and helps to reduce negative effects of stickiness. Apparently, this approach 
has not yet found an application in the context of Schwarzschild modelling.
If the initial conditions for orbit library are sampled at a small number of energy levels 
(which we do not encourage), one could average the contributions of all chaotic orbits with 
a given energy into one ``super-orbit'' which is then treated in modelling as any of the regular 
orbits \citep{MerrittFridman96}. The reason for this is that in 3D systems, all such orbits are 
parts of one interconnected (although not necessarily ``well-connected'') region in phase space%
\footnote{For models with figure rotation, this chaotic region is spatially unbounded for 
the values of Jacobi constant (which replaces energy as the classical integral of motion)
greater than the saddle point of the effective potential at Lagrange points. 
For this reason, the model of \citet{Hafner00} did not have any irregular 
orbits beyond corotation radius.}, so-called Arnold web \citep{LL92}. 

If necessary, one may enhance or reduce the relative fraction of chaotic orbits 
(or, in principle, any orbit family) by including an additional term in objective 
function, penalizing the use of such orbits \citep[e.g.][]{Siopis99}. 
However, \citet{VasilievAthanassoula12} demonstrated that decreasing the number of chaotic 
orbits in Schwarzschild model does not enhance the stability of the corresponding \Nbody model.

In our implementation, we have a linear or quadratic optimization problem, 
defined by a set of linear equations (\ref{eq_optim_delta}) and 
(\ref{eq_penalty_linear}) or (\ref{eq_penalty_tolerance_range}), optionally 
with additional quadratic penalty terms (\ref{eq_regularization_quadratic}) 
and/or penalties for given orbit families (e.g. tubes, chaotic orbits, etc.).
This optimization problem is solved by one of the available solvers, using 
an unified interface. Presently, three options are implemented: 
the BPMPD solver \citep{Meszaros99}, the CVXOPT library\footnote{http://cvxopt.org/} 
\citep{CVXOPT}, or the GLPK package\footnote{http://www.gnu.org/software/glpk/} 
(only for linear problems).
The typical time required to handle a model with $N_o={\cal O}(10^4), N_c={\cal O}(10^3)$
is within few minutes on a typical workstation, much less than the time needed to 
build the orbit library.

A Schwarzschild model may be converted to an \Nbody model by sampling each orbit in the 
solution by a number of points proportional to its weight in the model. 
For a collisionless simulation, the sampling scheme may be improved by using 
unequal mass particles \citep{ZempMSCM08, ZhangMagorrian08}, achieving better mass 
resolution and reducing two-body relaxation effects in the central parts of the model. 
The criteria for mass refinement may be based on energy, pericentre distance, 
or any other orbit parameter; the option for mass refinement is implemented in the code 
but has not been much explored.

\section{Tests for accuracy of potential approximations}  \label{sec_accuracy_potentials}

\begin{figure*} 
$$\includegraphics{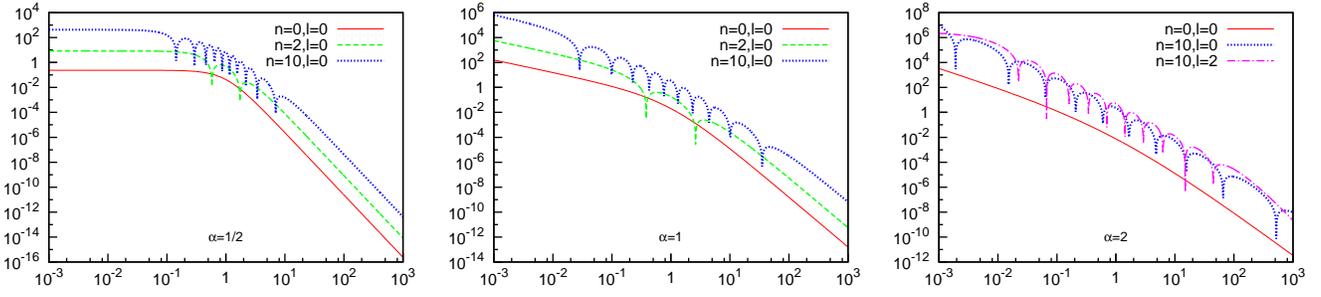} $$
\caption{ 
Radial part of density basis functions $|\rho_{nlm}(r)|$ (Eq.~\ref{eq_basis_rho}) for various 
$\alpha$: 
left -- $\alpha=1/2$, middle -- $\alpha=1$, right -- $\alpha=2$. 
Dips correspond to zero points of Gegenbauer polynomials, and the range of radii where 
the zeroes are found gives an idea of the range of radii where the approximation works well.
It is clear that increasing $\alpha$ leads to wider range of radii where the density can be 
well represented for a given order of expansion $\nmax$, at the expense of somewhat coarser
resolution at intermediate radii. 
The last plot shows also the non-spherical basis function of order $n=10, l=2$ -- it demonstrates 
that non-spherical components are well represented only in a narrower range of radii.
} \label{fig_basisfnc}
\end{figure*} 

\begin{figure*} 
$$\includegraphics[angle=-90]{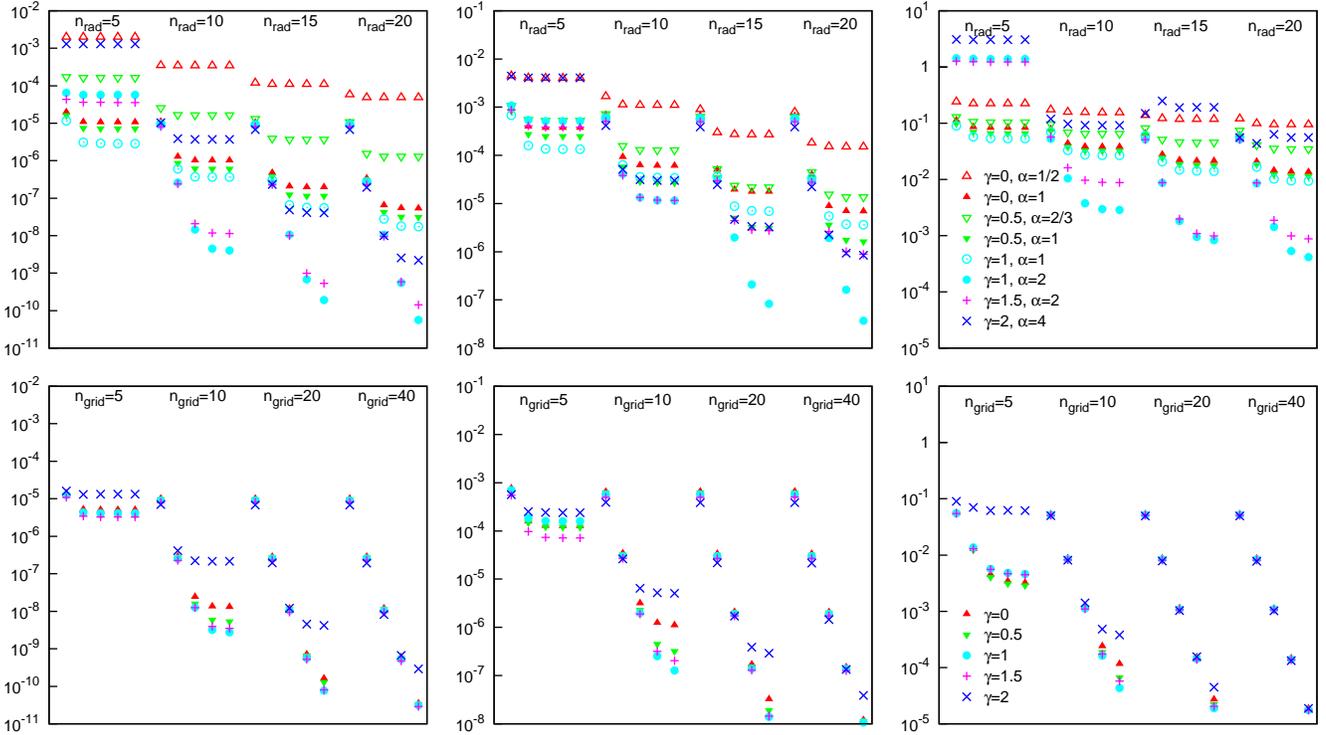} $$
\caption{ 
Integrated square relative errors for potential (left), force (center) and density (right) 
approximations of Dehnen models with basis-set (top) and Spline spherical-harmonic expansion 
(bottom).
Different symbols/colors represent models with different cusp slopes and $\alpha$ parameters 
in BSE. Four groups of points correspond to varying the number of radial coefficients $\nmax$ 
from 5 to 20 for BSE, and radial grid points for spline interpolation coefficients $\ngrid$ 
from 5 to 40; each group has 5 points for number of angular terms varying from $\lmax=2$ to 10 
(even numbers only).
\protect\\
\textbf{Top:} 
BSE models for $\gamma=0, 0.5$ and 1 are constructed using two different $\alpha$ parameters: 
first matches the inner cusp slope ($\alpha=1/(2-\gamma)$), second is a higher value of $\alpha$,
which, as seen from the figure, performs substantially better (open vs. filled triangles).
It is also clear that increasing $\lmax$ beyond 6 only makes a marginal improvement and only 
for some models, and $\nmax=15$ is, in general, sufficiently accurate. 
(The reason for non-monotonic behaviour of density error with $\lmax$ for the $\gamma=2$ model 
is a large relative deviation at $r\gtrsim 100$; at smaller radii approximations with larger 
number of terms are more accurate).\protect\\
\textbf{Bottom:}
Dehnen models with $\gamma\le 1$ are already well represented by spline approximations with 
$\ngrid=10$ radial points and $\lmax=6$ angular terms; 
steeper cusp slopes require somewhat larger $\ngrid$, with only $\gamma=2$ model benefiting 
from increasing this number to 40 (because of logarithmic divergence of potential at origin
it requires considerably smaller radius of the inner grid point). 
At large $\ngrid$, the number of angular terms is actually the factor that limits accuracy.
For almost all applications, $\ngrid=15-20$ and $\lmax=6-8$ will suffice and outperform BSE 
approximation.
} \label{fig_ise_from_exact}
\end{figure*} 

In this section we test the accuracy of three general-purpose potential approximations introduced 
above (basis-set, spline and $N$-body), from several different aspects. 
The first two potential expansions are smooth and should be able to represent analytically defined 
density profiles quite well, given a sufficient number of terms. 
All three are capable of approximating an arbitrary density model represented by a set of point 
masses. However, in this case there is a fundamental limit on the accuracy of approximation, 
set by discrete nature of underlying potential model; moreover, the optimal representation is 
achieved at some particular choice of parameters (order of expansion or $N$-body softening length), 
which needs to be determined: increasing the accuracy actually would only increase noise and not 
improve the approximation.

The accuracy of potential approximation is usually examined with the help of integral indicators 
such as ISE or ASE (integral/average square error: \citet{Merritt96, AthanassoulaFLB00}).
We compare the accuracy of representation of not only force, but also potential and density, 
and replace the absolute error with the relative one, since it better reflects the concept of 
accuracy, and allows to compare models with different underlying density profiles. 
Therefore, our measure of accuracy is defined by 
\begin{equation}  \label{eq_ISE}
\mathrm{ISE} = \int \rho(\boldsymbol{r}) 
  \left| 1-\frac{F_\mathrm{approx}(\boldsymbol{r})}{F_\mathrm{exact}(\boldsymbol{r})} \right|^2\,
  d\boldsymbol{r} 
\end{equation}

For all comparisons in this section we use a set of five Dehnen models with different cusp slopes
($\gamma=0, 0.5, 1, 1.5, 2$) and axis ratio of $1\!:\!q\!:\!p = 1\!:\!0.8\!:\!0.5$. 
The density profile is given by Eq.~(\ref{eq_Dehnen_profile}).
For $\gamma=1$ and $p=q=1$ this reduces to the spherical Hernquist profile, and for $\gamma=2$ -- 
to the Jaffe profile. 

We examine not only the integral error over the entire model, but also its variation with radius, 
to check which range of radii is well represented by the approximation, and what radii contribute 
the most to the integral error. 
For instance, a $\gamma=0$ model may have large relative errors in force approximation at small 
radii, because the true force tends to zero for $r\to0$ while the approximated one doesn't, 
but since the fraction of total mass at these small radii is negligible, it won't contribute to 
the total ISE. Nevertheless, one may argue that such a different behaviour of force at origin may 
substantially change the nature of orbits in the potential, for example, inducing more chaos as 
an orbit passes near the center.

Therefore, the next step is to compare the orbits in the true and approximated potential, which is 
done as follows. A set of $5000$ initial conditions, drawn uniformly from the corresponding 
density model at all radii (in the same way as for Schwarzschild modelling), is integrated in both 
the exact and approximate potentials for 100 dynamical times, and we compare several quantities on 
both per-orbit basis and on average. The properties of orbits to examine include the leading 
frequencies, LFCC diffusion rate $\Delta\omega$, Lyapunov exponent $\Lambda$, 
and the minimum squared angular momentum $L^2_\mathrm{min}$ (which distinguishes between 
centrophilic and centrophobic orbits).
As explained in Section~\ref{sec_freq_analysis}, $\Delta\omega$ is not a strictly defined 
quantity, and so we could not expect it to match perfectly at the level of individual orbits; 
however, the scatter in $\log(\Delta\omega_\mathrm{approx}/\Delta\omega_\mathrm{exact})$ shouldn't 
be much larger than the expected $0.3-0.5$ dex for a smooth potential, and there should not be 
any systematic shift towards higher (or lower) $\Delta\omega$. Similar considerations apply to 
$\Lambda$.

\subsection{Approximating a smooth potential model with basis set or splines}  \label{sec_accuracy_vs_exact}

\begin{figure*} 
$$\includegraphics[angle=-90]{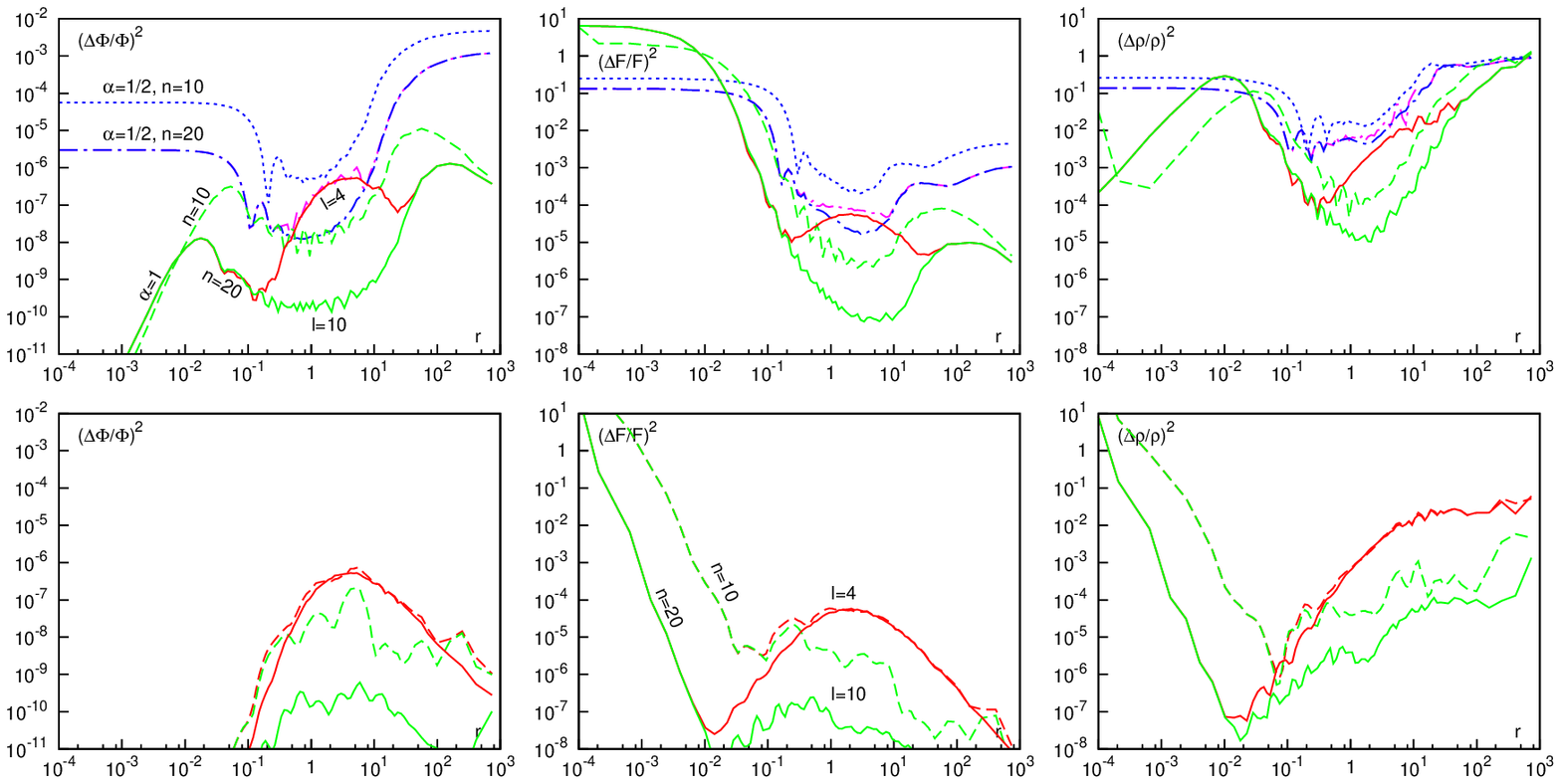} $$
\caption{
Relative squared errors for potential (left), force (center) and density (right) approximations 
of $\gamma=0$ Dehnen model, as functions of radius.  \protect\\
\textbf{Top:} BSE with $\alpha=1/2$ (blue dotted line for $\nmax=10$, blue dot-dashed for 
$\nmax=20,\lmax=10$, purple dot-dashed for $\nmax=20,\lmax=4$), and 
with $\alpha=1$ (green dashed -- $\nmax=10, \lmax=10$, green solid -- $\nmax=20,\lmax=10$,
red solid -- $\nmax=20,\lmax=4$.
This panel illustrates that BSE with larger $\alpha$ have larger useful radial range, even if 
the inner cusp slope doesn't match that of the underlying model.
Increasing the number of radial functions does extend the range of radii in which approximation
works well (trough-like shape of the bottom curve at intermediate radii, $r\sim 10^{-1}-10^1$);
for a fixed $\nmax$, increasing the order of angular expansion improves approximation at 
these intermediate radii (compare solid curves for $\lmax=4$ and 10), but only until the 
``bottom of the trough'' is reached. At small or large radii increasing $\lmax$ has no effect.
It is the integral over these intermediate radii which mainly contributes to the ISE values 
of Fig.~\ref{fig_ise_from_exact}, but large relative errors at small or large radii may be hidden 
in that integral characteristic. \protect\\
\textbf{Bottom:} Spline approximations with $\ngrid=10$ (dashed) and $\ngrid=20$ (solid),
top (red) is for $\lmax=4$, bottom (green) -- for $\lmax=10$. 
Clearly this expansion performs much better overall than BSE, and continues to improve with 
increasing $\lmax$ for given $\ngrid$, saturating at smaller errors and in larger radial range.
} \label{fig_error_radial_profile}
\end{figure*} 

First we examine the accuracy of basis-set representation of a smooth potential.
The basis sets used in the literature are complete in the sense that they may approximate any 
well-behaved potential-density pair, given a sufficient number of terms. However, to be effective 
with a small number of terms, their lowest-order term should resemble the underlying profile as 
close as possible. For example, the Clutton-Brock basis set is not very suitable for representing 
density profiles with central cusps. 

The \citet{Zhao96a} basis set implemented in \SMILE is based on the two-power density profile 
with the inner and outer slopes equal to $-2+1/\alpha$ and $-3-1/\alpha$ correspondingly 
(section~\ref{sec_app_bse}), 
where the parameter $\alpha\ge 1/2$ may be chosen to give the highest possible accuracy for 
a given density profile. One might think that, for example, a better approximation to a model 
with finite central density is obtained with a value of $\alpha=1/2$ (corresponding to the 
cored Plummer profile as the zero-order term), but it turns out that matching the cusp slope 
is not necessarily the best idea.
More important is the range of radii in which the basis-set approximation is reasonably good,
which depends both on the maximal order of expansion $\nmax$ and on $\alpha$: 
higher $\alpha$ give a greater range because the break in density basis functions is more 
extended, and because their zeroes cover larger range of radii (Fig.~\ref{fig_basisfnc}).

For each value of $\gamma$ we constructed a series of BSE approximations with the number of 
radial terms $\nmax$ varying from 5 to 20 (in steps of 5) and the angular expansion order $\lmax$ 
from 2 to 10 (even values only). Fig.~\ref{fig_ise_from_exact}, top panel, shows the integrated 
relative squared errors in potential, force and density approximations. A general trend is that 
increasing $\nmax$ always makes errors smaller, and increasing $\lmax$ improves the approximation 
up to $\lmax=6$, after which there is no appreciable difference in most cases. 
This may be understood as ``saturation'' of the approximation accuracy in the range of radii 
which contributes the most to the integral quantity. 

An interesting result is that for weak-cusp models, increasing $\alpha$ above the value 
corresponding to the inner cusp slope actually makes the approximation much better. 
The reason is just a greater useful radial range of higher-$\alpha$ basis sets, as exemplified 
in Fig.~\ref{fig_error_radial_profile}, top, for $\gamma=0$ and $\alpha=1/2$ (Clutton-Brock) 
vs. $\alpha=1$ (Hernquist-Ostriker) basis sets. The latter clearly performs better at larger 
range of radii, and the improvement of error saturates at larger values of $\lmax$ 
(for the former, there is no practical difference beyond $\lmax=4$). 
Similarly, even for $\gamma=1$ Dehnen model which is traditionally represented with $\alpha=1$ 
Hernquist-Ostriker basis set, the $\alpha=2$ approximation actually works much better, both in 
the integral sense and in the range of radii for which relative error is small and improves 
with increasing $\lmax$ (``bottom of the trough'' depicted on the above figure). 
However, for even greater $\gamma$ it doesn't make sense to increase $\alpha$ beyond 
the value corresponding to the inner cusp slope; that is, for $\gamma=1.5$ model $\alpha=2$ 
is the best choice. The case $\gamma=2$ is particularly difficult, since the potential diverges 
at origin, and formally $\alpha\to \infty$; we restrict this parameter to be $\le 4$ for the 
reason that the magnitude of coefficients rapidly increases with $\alpha$ and $l$, and roundoff
errors become intolerable.

\begin{figure} 
$$\includegraphics{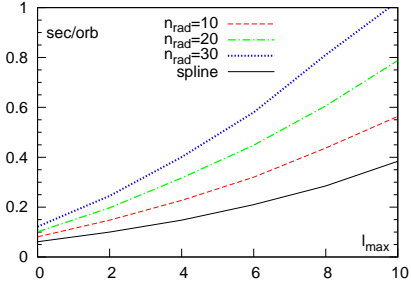} $$
\caption{
Wall-clock integration time of a single orbit on a typical workstation (for 100 orbital periods), 
depending on the number of terms in BSE and spline approximations. 
Horizontal axis is the order of angular expansion, with the number of terms being 
$\frac{1}{2}(\frac{\lmax}2+1)(\frac{\lmax}2+2)$. 
Solid line is for Spline with $\ngrid=40$ (it almost does not depend on number of radial points),
other lines from bottom to top are for BSE with $\nmax=10$, 20 and 30 radial terms. 
For comparison, integration time for an exact triaxial $\gamma=1$ Dehnen potential is 
$\sim 1.5$~sec/orb.
} \label{fig_integration_time}
\end{figure}

In the case of Spline expansion, there is an additional freedom of choice of grid nodes, 
either to get a higher resolution (more frequently spaced nodes) at the intermediate radii 
where the bulk of integrated error comes from, or to achieve a better approximation at small 
or large radii. We find that exponentially spaced nodes are a good way to afford a large 
dynamic range in radius with relatively few nodes ($\ngrid \sim 10-20$ for 
$r_\mathrm{out}/r_\mathrm{in}\gtrsim 10^4$), 
so that the adjacent nodes differ by a factor of $1.5-3$ in radius. Under these conditions, 
the accuracy at intermediate radii is mostly limited by order of angular expansion, 
at least for $\ngrid\ge 10$ and $\lmax\lesssim 8$; only the steepest cusp slopes require 
more than 20 nodes to achieve really small errors. Overall, the spline expansion outperforms 
BSE for comparable number of coefficients, both in terms of the radial range in which errors 
are small, and in integral characteristics such as ISE 
(Figs.~\ref{fig_ise_from_exact} and \ref{fig_error_radial_profile}, bottom panels).
In terms of computational efficiency of orbit integration, spline expansion is also faster 
than BSE and its performance is almost independent of the number of radial nodes (at least 
up to $\ngrid=40$). Both are substantially, by a factor of few, faster than the exact Dehnen 
potential (Fig.~\ref{fig_integration_time}).

Finally, we compare the properties of orbits integrated in exact and approximated potentials,
to address the question how much the relative errors in potential and especially force 
affect the dynamics. In particular, many of the approximate models have asymptotic behaviour 
of force at small radii which is different from the exact model (in particular, BSE with 
the parameter $\alpha$ not matching the inner cusp slope). It is not obvious to which extent 
these deviations actually matter, without comparing the actual orbits. 
Even if properties of individual orbits do not strictly match between exact and approximate 
potentials, we want the ensemble of orbits to exhibit similar characteristics 
(e.g. distribution in $\Delta\omega$, $\Lambda$, number of centrophilic orbits, etc.).

These studies basically confirm the conclusions of the above discussion. 
For BSE approximations, almost any model with $\nmax \ge 10$ and $\lmax \ge 6$ is 
close enough to the exact potential. For given $\gamma$, models with higher $\alpha$ 
are better approximations (that is, for $\gamma=0$ the case $\alpha=1/2$ performs much worse 
than any other model, and $\alpha=1$ is already good; for $1/2 \le \gamma \le 3/2$ the models 
with $\alpha=2$ are preferred). 
Comparing properties of individual orbits, we find that frequencies $\omega$, orbit shape 
(measured by diagonal values of the inertia tensor), and values of minimum squared angular 
momentum $L^2_\mathrm{min}$ are recovered to within few percents, and $\log \Delta\omega$ 
typically has scatter of $0.5-1$ (this is the only quantity which improves steadily with 
increasing $\nmax$.
For the entire orbit ensembles, distribution of orbits in $\Delta\omega$ and in $\Lambda$ 
is very close to the one from exact models for all cases with $\nmax\ge 10,\lmax\ge 4$ 
(with the above mentioned exception of $\gamma=0,\alpha=1/2$ model). 
The fraction of centrophilic orbits is determined with $\sim 5\%$ accuracy (and doesn't 
further improve with increasing the precision), and the number of orbits with $\Lambda=0$ 
matches to within $1\%$. 
For spline expansion, conclusions are similar; $\ngrid=20$ was sufficient for all models 
except $\gamma=2$, for which 40 nodes did show improvement over 20; and there is no 
substantial change after $\lmax=6-8$.

\begin{figure} 
$$\includegraphics{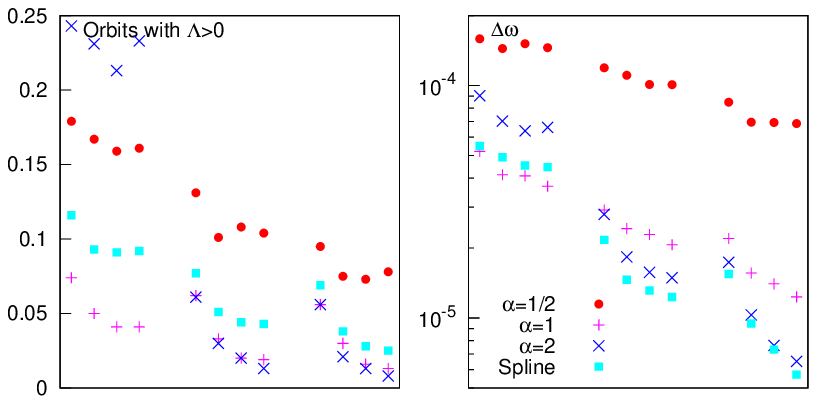} $$
\caption{
\textbf{Left: } Fraction of chaotic orbits in BSE and Spline approximations of the Perfect 
Ellipsoid potential (with $q=0.8,p=0.5$), detected by computing Lyapunov exponent over 100 
dynamical times.\protect\\
\textbf{Right: } Mean value of frequency diffusion rate $\Delta\omega$. \protect\\
Three groups correspond to $\nmax=10,15,20$ for BSE and $\ngrid=10,20,40$ for Spline; 
points in each group have $\lmax=4,6,8,10$.
The exact potential is integrable, therefore the lower is the number of orbits with $\Lambda>0$
or the value of $\Delta\omega$, the better is the approximation. 
It is clear that while all approximations do perform better with increasing number of terms, 
some do it much faster: generally, $\alpha=2$ BSE expansion wins the race even though its 
behaviour at origin is very different from the flat core of the Perfect Ellipsoid.
} \label{fig_perfect_ellipsoid}
\end{figure} 

Another test of the same kind is a study of orbital properties of BSE/Spline approximations of 
the Perfect Ellipsoid \citep{Kuzmin56, deZeeuw85}, which is a fully integrable triaxial potential 
corresponding to the density profile $\rho(\tilde r) = (\pi^2pq)^{-1}(1+\tilde r^2)^{-2}$. 
Since all orbits in the exact potential should be regular, we may easily assess the quality 
of approximation by counting the number of chaotic orbits. 
Fig.~\ref{fig_perfect_ellipsoid} shows that indeed BSE approximations with higher $\alpha$ 
parameters are much better at representing the potential, despite that their asymptotic behaviour 
of force at origin is different from the exact potential. 

Overall conclusion from this section is that BSE and spline expansions with a sufficient number 
of terms are good approximations to Dehnen models with all values of $\gamma\in [0..2]$. 
For BSE, the parameter $\alpha=2$ gives best results for $1/2\le \gamma \le 3/2$, with 
$\gamma=0$ and 2 requiring $\alpha=1$ and 4, correspondingly; $\nmax=10-15$ is sufficient.
The order of angular expansion $\lmax=6-8$ is enough for moderately flattened systems 
considered in this section, but may need to be increased for highly flattened, disky models.

\subsection{Basis-set and spline representation of a discrete particle set}  \label{sec_accuracy_from_finite_N}

A rather different case is when the coefficients of potential expansion are evaluated from 
a set of point masses, for example to study orbital properties of an \Nbody system.
As is general for this kind of problems, the approximation error is composed of two terms.
The \textit{bias} is the deviation of approximated potential from the presumable ``intrinsic'' 
smooth potential, which is the continuum limit of the \Nbody system, and was explored in the 
previous section: increasing number of terms never makes it worse, although may not improve 
substantially after a certain threshold is reached.
The \textit{variance} is the discreteness noise associated with finite number of particles, and 
it actually increases with the order of expansion. Therefore, a balance between these two terms 
is achieved at some optimal choice of the number of coefficients \citep[e.g][]{Weinberg96}.
This conclusion is well-known in the context of choice of optimal softening length in 
collisionless \Nbody simulations \citep{Merritt96, AthanassoulaFLB00, Dehnen01}, 
and will also be reiterated in the following section in application to the tree-code potential. 
Here we show that a similar effect arises in the smooth BSE and Spline representations 
of a discrete particle set.

\begin{figure} 
$$\includegraphics{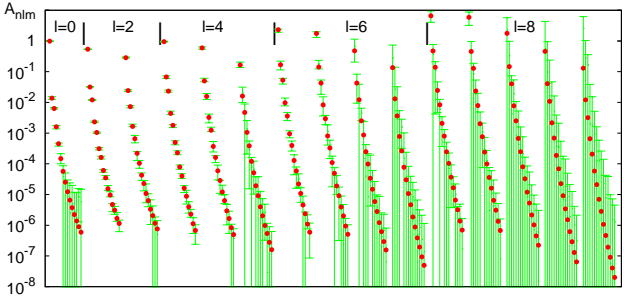} $$
\caption{
Coefficients of a $\alpha=1$ BSE approximation of a triaxial $\gamma=0.5$ Dehnen model.
Groups of 15 points represent radial coefficients at a given angular harmonic $l,m$; 
values of $l$ increase from 0 to 8 (even only) and $m$ from 0 to $l$ between groups.
Points are the coefficients from analytical density profile, error bars are average over
10 realizations of $10^5$ points; error bar extending to zero means that value of this 
coefficient is noise-dominated in the discrete realizations. 
} \label{fig_bse_coef_scatter}
\end{figure} 

\begin{figure} 
$$\includegraphics{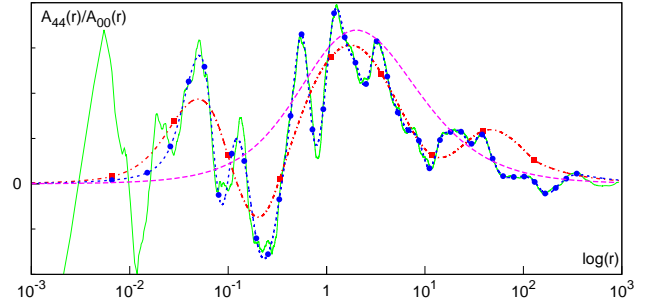} $$
\caption{
Spline approximation of the radial variation of $l=4,m=4$ spherical-harmonic coefficient 
(normalized to $l=0$ coefficient) for a $10^5$ particle $\gamma=1$ Dehnen model.
Green solid line -- coefficients evaluated at particle positions; 
red dash-dotted and blue dashed lines -- spline approximations with $\ngrid=10$ and $40$,
correspondingly; grid nodes are marked with beads (square for $\ngrid=10$, round for 40).
Purple dashed line -- coefficients computed from exact density profile. \protect\\
It is clear that increasing the number of grid points may lead to an almost exact representation
of $A_{lm}(r)$ for a given \Nbody snapshot, but it does not converge to the ``true'' coefficient 
for the exact underlying density model, and actually deviates from it more as we increase $\ngrid$.
} \label{fig_spline_coefs_fitting}
\end{figure} 

An illustration of the effect of variance is provided by the following exercise. 
We generate several realizations of the same triaxial density profile, compute expansion 
coefficients for each one and calculate their average values and dispersions, comparing 
to the coefficients evaluated from the analytical density model. 
Fig.~\ref{fig_bse_coef_scatter} demonstrates that all BSE coefficients with sufficiently 
high indices $n,l,m$ are dominated by noise. Somewhat surprising is the small number of 
``usable'' terms -- just a few dozen even for a $N=10^5$ particle model. The range of 
significant terms depends on the number of particles and on the details of density 
distribution and $\alpha$ parameter in BSE, but in general, angular terms beyond 
$l=6 (8)$ for $N=10^5 (10^6)$ model are unreliable, with only a few first $n$ and $m$ 
terms at that value of $l$ are significant. 

For Spline potential, the situation is similar, but instead of variation of coefficients 
with $n$, we follow their variation in radius for a given $l,m$. 
Noise limits the useful order of angular expansion especially at small radii, where 
the interior mass is represented by just a small number of points; 
for intermediate to larger radii the coefficients are reliable for a somewhat higher 
angular order (e.g. up to $l=8,m\le 2$ for $N=10^5$ model).
Fig.~\ref{fig_spline_coefs_fitting} shows the radial variation of the $l=4,m=4$ 
spherical-harmonic coefficient for a particular $10^5$ particle model, together with spline 
approximations with $\ngrid=10$ and 40 points, compared to the coefficient from 
exact density profile. It is clear that in this case, increasing number of nodes may 
make the spline approximation match the radial dependence of this coefficient almost perfectly,
but it turns out to fit mostly discreteness noise rather than true behaviour of this 
harmonic. While some regularization techniques may be applied to achieve balance between
approximation accuracy and spline smoothness \citep[e.g.][]{GreenSilverman94}, it may be 
easier just to keep the number of radial grid nodes small enough ($10-15$ in this case).

We explore the accuracy of approximation for the same five Dehnen profiles as in the previous
section, but now initializing coefficients from $N=10^5$ and $10^6$ particle realizations 
of corresponding models. Again we compare the ISE indicator and the range of radii for which 
the error is tolerable, as well as the orbital properties.

Not surprisingly, it turns out that the ISE in potential, force and density approximations 
cannot be reduced below a certain level, which is roughly $10^{-6}, 10^{-4}$ and $10^{-2}$
correspondingly, for a $10^5$ particle model, and somewhat lower for $N=10^6$ model.
There is almost no improvement in integrated error after $l=4,n=10$ for $N=10^5$.
In terms of orbit properties, however, higher $\lmax$ and $\nmax$ actually make 
things worse, in the sense that orbits become more irregular compared to the exact potential.
If the number of coefficients is too low ($\lmax=2$ or $\nmax=5$), the model is too far 
from the exact one, i.e. the difference in orbit properties is dominated by \textit{bias}:
there are fewer chaotic orbits and their properties are noticeably different from the 
exact potential. $\nmax=10,\lmax=4$ seems to be the optimal choice for a $N=10^5$ model
and $\nmax=15,\lmax=6$ is comparably good as the previous choice for a $N=10^6$ case.
Taking the expansion to higher orders introduces substantial bias in orbit properties:
number of chaotic ($\Lambda>0$) and centrophilic orbits is increased, and $\Delta\omega$ 
is noticeably shifted to higher values. 
Other properties of orbits are recovered quite well and are insensitive to the choice 
of expansion order: the mean-square difference in orbital frequencies is $\lesssim 5\%$, 
and the orbit shape is accurate to within 2\%.
Also worth mentioning is that here the results from ISE and orbit studies disagree: 
while the former indicator is still improving or at least constant with increasing 
$\lmax$ and $\nmax$, the divergence in chaotic properties of orbits is noticeably 
increasing beyond the optimal choice of these parameters. 
Actually, the integral error is often accumulated at small radii, where the enclosed 
number of particles is small (${\cal O}(10)$, with relative force and density errors
$\gtrsim {\cal O}(10^{-1})$), but apparently the influence of these deviations on the 
gross dynamics is not very strong; more important are the high-frequency fluctuations 
at all radii, caused by higher harmonics whose amplitudes are pumped up by noise.

To summarize, representing a discrete \Nbody system by a smooth potential approximation 
requires a judicious choice of order of expansion, which balances the contribution 
of bias and variance to the approximation error. A rough estimate of how many terms 
should be retained can be obtained by evaluating expansion coefficients for several 
snapshots of the system, and retaining only those coefficients which do not have scatter 
greater than their magnitude. (Alternatively, one may wish to average over several 
snapshots to get a lower effective discreteness limit, but still the truncation order 
should be compatible with the data).
If multiple snapshots are not available, a rule of thumb would be to leave terms 
up to $\lmax=4 (6)$ and $\nmax \sim 10-15$ for $N=10^5 (10^6)$ particle model.

\subsection{Optimal softening for an $N$-body potential}  \label{sec_accuracy_treecode}

In this section we explore the accuracy of the tree-code potential solver, 
which evaluates the potential directly from a set of $N$ point masses, without 
approximating it as a basis-set or spline spherical-harmonic expansion. 
There are two tunable parameters in this potential solver, the softening length $\epsilon$ 
and the opening angle $\theta$ of the tree-code. 

Since our intention is to use the \Nbody snapshot as a discrete representation of a smooth 
mass distribution, it makes sense to introduce some sort of softening, for the same reasons 
as for the construction of smoothing splines from a discrete point mass set: 
the frozen \Nbody potential typically does not mean to represent this particular collection 
of particles, but rather samples a given smooth potential with a discrete set of points. 
Therefore, the error in this representation arises from two constituents: \textit{variance} 
coming from fluctuations in particle positions, and \textit{bias} from replacing Newtonian 
potential with a softened one. These two errors depend in the opposite way on $\epsilon$, 
so there exists a formally optimal value of softening length associated with a given mass 
distribution and $N$ \citep{Merritt96, AthanassoulaFLB00}.
Introduction of a softening length $\epsilon$ and modification of the Newtonian potential 
of particles at radii $\lesssim \epsilon$ is essentially equivalent to considering the 
gravitational field created by a smoothed density distribution (see \citet{Barnes12} 
for a discussion), therefore the problem of determining the optimal softening length is 
reduced to the problem of density estimation from a discrete point set. 
It is reasonable that the choice of softening length should depend on the local 
density \citep{Dehnen01}, although the optimal value of $\epsilon$ for the potential, 
acceleration and density estimation may differ by a factor of few. 
Collisionless \Nbody codes with spatially variable and time-adaptive $\epsilon$ are 
not widespread due to additional complications arising from the necessity of modifying 
the equations of motion to account for time-varying softening (\citet{IannuzziDolag11} 
describe one such implementation), but there is no reason why such spatially adaptive 
softening should not be used in a frozen \Nbody potential solver.

Another source of error in force computation in the tree-code comes from the tree 
algorithm itself, which replaces many particles with single tree cells. 
The additional error in force coming from this approximation depends on the cell 
opening angle $\theta$ and is generally rather low compared to the noise due to 
the discreteness of mass distribution \citep{HernquistHM93}. 
Indeed, the tests below demonstrate that the error in the force and, 
to a lesser extent, potential approximation is rather insensitive to the accuracy 
of force calculation itself (opening angle $\theta$ and the use of quadrupole versus 
monopole terms), and is limited by discreteness noise. However, the error in 
calculation of an orbit does depend on the accuracy of tree-code (and on the integration 
timestep). To keep the errors in energy conservation at acceptable level, one should 
therefore appropriately choose the parameters of the tree-code.

The first set of tests is therefore dedicated to the determination of optimal softening 
parameters for a static approximation of a known potential by its discrete point mass 
representation.  
There are many choices for the functional form of softening. We define the softening kernel 
$\hat\rho(x)$ so that the density associated with a single particle of mass $m_i$ at location 
$\boldsymbol{r}_i$ with a softening length $\epsilon_i$ is given by 
$m_i \epsilon_i^{-3}\hat\rho(|\boldsymbol{r}-\boldsymbol{r}_i|/\epsilon_i)$. 
The simplest form, widely used in \Nbody simulations, is the Plummer softening, given by 
$\hat\rho(x)\equiv (3/4\pi)(1+x^2)^{-5/2}$; another choice, often employed in the non-parametric 
density estimation, is the Ferrers $n=1$ (also called Epanechnikov) kernel: 
$\hat\rho(x)\equiv (15/8\pi) (1-x^2)$ for $x<1$ and zero for $x\ge 1$.
The adaptive softening length for each particle of mass $m_i$ is taken to be 
\begin{equation}
\epsilon_i = \varepsilon\, (m_i/\overline m)^{1/2} n_i^{-1/3} ,
\end{equation}
where $\overline m\equiv M_\mathrm{total}/N$ is the average particle mass, $n_i$ is the 
estimate of the local number density and $\varepsilon$ is the global softening parameter. 
This local variation of softening length results in substantially lower bias than a single 
$\epsilon$ for all particles \citep{Dehnen01}. 
The tests below are done for equal values of $m_i$ and seek to determine the optimal value 
of $\varepsilon$, which in this case is the coefficient of proportionality between the 
softening length and the local mean interparticle distance.

\begin{figure} 
$$\includegraphics{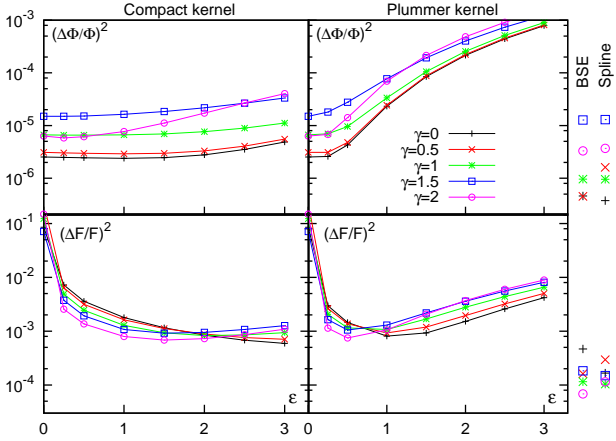} $$
\caption{
Integrated square relative errors for potential (top row) and force (bottom row), 
evaluated for a $N=10^5$ particle potential using the \Nbody tree-code potential
with spatially-adaptive softening length, proportional to mean inter-particle distance 
with a coefficient $\varepsilon$. The dependence of the error on the softening 
parameter is plotted for five Dehnen models with $\gamma$ from 0 to 2, using 
the compact Ferrers kernel (left) and Plummer softening (right). 
The curves demonstrate the bias-variance tradeoff: increasing $\varepsilon$ 
decreases the random variance of potential but increases the bias, so there is 
an optimal choice for $\epsilon$ that minimized the total error.
It is clear that the compact kernel is better suited for the adaptive softening, 
since it has a fairly shallow dependence on the softening parameter, while for 
the Plummer kernel the bias rapidly becomes large with increasing $\varepsilon$.
For comparison, in the right panel we also plot the errors obtained using the BSE 
or Spline expansions initialized from the same \Nbody snapshot 
(using $\nmax=20$ radial and $\lmax=6$ angular terms), 
which are several times lower than achievable with the discrete point mass potential 
even with the right amount of smoothing.
} \label{fig_ise_from_treecode}
\end{figure} 

We consider the same five Dehnen models as in the previous sections.
Each model is sampled with $10^5$ or $10^6$ equal point masses, and the relative square error 
in potential and force approximation is plotted as a function of radius and the integrated value 
(ISE) is computed, as a function of the softening parameter $\varepsilon$.
Fig.~\ref{fig_ise_from_treecode} shows that for the Plummer softening there is a well-defined 
minimum in ISE of force approximation at $\varepsilon \sim 0.5-1$, while for the potential 
approximation the no-softening case gives formally the best result. This is because larger 
values of $\varepsilon$ substantially increase the bias, since the Plummer softening modifies 
the potential at all radii and that dominates the approximation error. For the compact softening 
kernel the dependence of ISE on $\varepsilon$ is much weaker; values of $\varepsilon \sim 1-2$ 
give near-optimal results in most cases, with steeper cusps requiring smaller $\varepsilon$.
This is in agreement with \citet{Dehnen01} who demonstrated that Plummer softening is inferior 
to compact kernels for various reasons. There is almost no dependence on the accuracy of the 
tree-code algorithm for the force estimation, while for the potential estimation the inclusion 
of quadrupole moments and a smaller opening angle $\theta$ gives a better result.

Interestingly, but perhaps not surprisingly, approximation of an \Nbody snapshot with a BSE 
or Spline potential (section~\ref{sec_accuracy_from_finite_N}) gives a substantially better 
result in terms of ISE, not even speaking of efficiency of orbit integration. 
This suggests that if one wants to analyze the orbital structure of an \Nbody system, such 
expansion methods \citep[e.g.][]{HoffmanCDH10} are preferred over a frozen \Nbody potential 
representation \citep[e.g.][]{Valluri10}. Not only they allow for a better accuracy of 
the force approximation, but also enable to compute Lyapunov exponents and even the frequency 
diffusion rate with a greater precision.

Secondly, to study how the granularity of the potential affects the chaotic properties of orbits, 
and to determine the parameters of tree-code that do not compromise the accuracy of orbit 
integration, we take an \Nbody realization of a triaxial Perfect Ellipsoid (with axis ratio 
$1\!:\!0.8\!:\!0.5$) which is known to be an integrable potential, and calculate $10^3$ orbits 
covering the entire range of energies. The chaoticity of orbits is measured by the mean value of 
frequency diffusion rate (FDR) $\overline{\Delta\omega}$, which should be as low as possible 
(Fig.~\ref{fig_perfect_ellipsoid} shows that it is around $10^{-5}$ for a BSE or Spline 
potential constructed from an analytic density profile). 
The force computed by the tree algorithm is not a continuous function of position, it has jumps 
when a cell opening criterion is triggered for any group of particles. As a result, energy 
of an orbit is not conserved exactly, no matter how small timesteps are taken, and this limits 
the accuracy of frequency determination. 
Fig.~7 in \citet{VasilievAthanassoula12} suggests that as long as the energy conservation error is 
kept below some threshold, it does not affect the values of FDR for orbits in the \Nbody potential. 
To estimate this threshold and the necessary parameters of tree-code, we run the orbit integration 
for various values of opening angle $\theta=1,0.7,0.5,0.4$ and softening parameters $\varepsilon$.
We found that an opening angle $\theta \le 0.5$ is necessary to achieve a reasonable energy 
conservation error of $\sim 10^{-3}$, which is well below the average value of 
$\Delta\omega = 10^{-2.1} (10^{-2.4})$ for a $N=10^5 (10^6)$ particle model. 
Better results are obtained with larger softening parameter (up to $\varepsilon=3$).
For comparison, orbits integrated in BSE and Spline potentials initialized from the same \Nbody 
snapshots had $\overline{\Delta\omega} \sim 10^{-4}$, and the lowest value was obtained 
for $\lmax=4(6)$ angular terms for $N=10^5(10^6)$, with little dependence on the number of 
radial terms $\nmax$. This is still higher than for an analytic density profile, but way better 
than for the discrete potential, and allows to perform a meaningful detection of chaotic orbits 
using a conservative threshold of $\Delta\omega_\mathrm{ch}=10^{-3}$.

\section{Tests of variants of Schwarzschild models}  \label{sec_schw_modelling_tests}

Here we present comparison of the three variants of Schwarzschild modelling 
presented in Section~\ref{sec_schw_variants}. 
We use two test cases: the first one is a triaxial $\gamma=1$ Dehnen model with 
a central black hole, and the second one is a mildly cusped strongly triaxial model 
created by a cold collapse, which does not have an \textit{a priori} known analytical 
density profile.

\subsection{Triaxial Dehnen model with a central black hole}

The first test case is the $\gamma=1$ Dehnen model with axes ratio 
$x\!:\!y\!:\!z=1\!:\!0.8\!:\!0.5$, with a total mass of unity, and a central point mass 
$M_\bullet=0.01$ representing a galactic supermassive black hole (BH).
It is well established \citep[e.g.][]{ValluriMerritt98} that such a model with 
a relatively large central point mass has a large fraction of chaotic orbits at radii 
larger than a few BH influence radius $r_h$ (defined as the radius containing stellar mass 
equal to $2M_\bullet$). 

Earlier studies \citep{GerhardBinney85, MerrittQuinlan98} suggested that when 
$M_\bullet \gtrsim 2\%$ of the total mass of the galaxy, its effect on the centrophilic 
orbits is strong enough to destroy triaxiality in just a few crossing times.
Subsequent works, however, presented examples of a long-lived triaxial models with 
comparably large central masses: for instance, \citet{HolleyMSHN02} considered a 
moderately triaxial $\gamma=1$ Dehnen model with $M_\bullet=0.01$ in which the loss 
of triaxiality was confined to the inner 2\% of particles, while \citet{KalapotharakosVC04} 
found that such a black hole mass could drive a model towards axisymmetry in a time 
comparable to Hubble time, and \citet{PoonMerritt04} demonstrated that truncated 
power-law models of inner parts of a galaxy (enclosing total mass of a few times 
$M_\bullet$) can remain triaxial even with a large fraction ($\gtrsim 50\%$) 
of chaotic orbits. 

The goal of this test case is not to present a comprehensive study of evolution 
of triaxial cuspy models with black holes, but to show that at least in some cases 
the Schwarzschild method can be used to construct a relatively stable model.
We use the three variants of Schwarzschild modelling techniques considered above 
(classical, basis-set and Spline spherical-harmonic expansion, hereafter labelled 
as C, B and S), with $N_r=30$ radial shells or radial basis functions and 
27 (28) angular constraints for C (B, S) models, giving $N_c\simeq 800$ total constraints 
for $N_o=4\times 10^4$ orbits. All three models were solved with quadratic optimization 
routine and were required to keep velocity isotropy at all radii.
The fraction of chaotic orbits was around 30\%. 
Then a $N=10^5$ \Nbody model was created from each Schwarzschild model and evolved 
for $T=100$ time units with the direct, hardware-accelerated \Nbody integration code 
$\phi$GRAPEch \citep{HarfstGMM08}, which uses algorithmic regularization to improve 
the accuracy of integration of trajectories that come close to the black hole.
The softening length was set to zero and the accuracy parameter of the Hermite 
integrator was set to $\eta=0.01$; total energy was conserved to a relative accuracy 
better than $\sim 10^{-4}$.

\begin{figure} 
$$\includegraphics{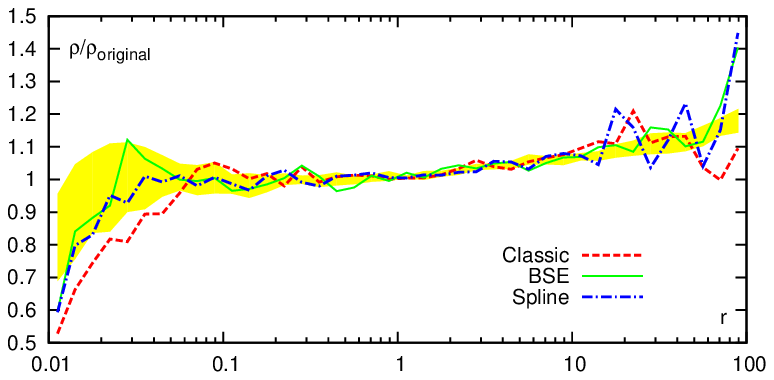} $$
$$\includegraphics{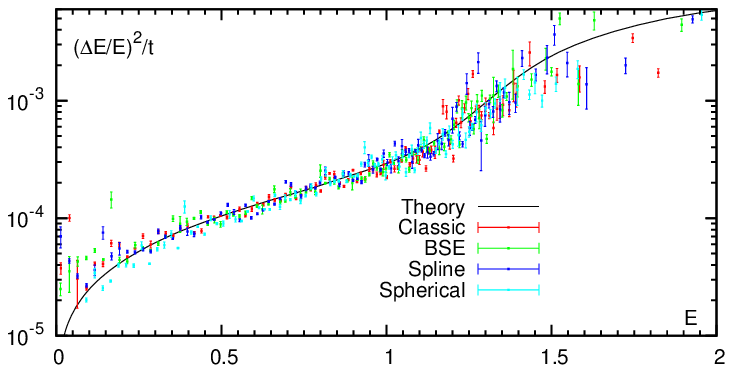} $$
\caption{
Three variants of Schwarzschild models for a triaxial $\gamma=1$ Dehnen model with 
a central black hole $M_\bullet=0.01$. \protect\\
\textbf{Top panel:} density normalized to the expected value, as a function of radius. 
Shaded region shows typical uncertainty region for a random realization of a $10^5$ 
particle model (in which particle positions are distributed according to the analytical
profile). \protect\\
\textbf{Bottom panel:} energy diffusion rate (relative mean-square change of particle 
energy per unit time, as a function of initial energy).
Solid line shows theoretically computed value, points with error bars are the results 
from the simulation. Particles are binned into 100 bins of unequal size, with the most 
bound bins each containing $\sim 10$ particles; error bars represent the uncertainties 
in the estimate of the slope of energy growth with time. Different colors are for 
the three models, but there is no systematic difference between them. A spherically 
symmetric model with the same azimuthally-\-averaged density profile, created by 
Eddington inversion formula, is also plotted for comparison; it demonstrates 
the same energy diffusion rate as the triaxial Schwarzschild models.
} \label{fig_d1bh001_dens_energydiff}
\end{figure} 

\begin{figure*} 
$$\includegraphics{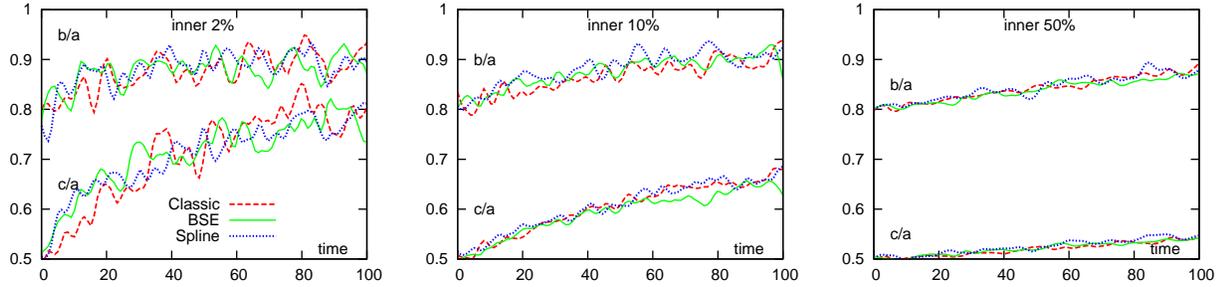} $$
\caption{
The shape evolution of the three models from the previous figure. Three panels show 
the axis ratios of inertia tensor containing 2, 10 and 50\% of mass (excluding the 
black hole), as functions of time. The evolution towards more spherical shape occurs 
most rapidly at small radii, but does not lead to a complete loss of triaxiality 
even in the centre. The three variants of Schwarzschild models demonstrate similar
amounts of evolution.
} \label{fig_d1bh001_shape}
\end{figure*} 

To check how close the models are to equilibrium, we used a number of indicators. 
First the density profiles of the \Nbody models were compared to that of the analytical 
mass profile used to create Schwarzschild model. Fig.~\ref{fig_d1bh001_dens_energydiff}, 
top panel, shows the ratio of densities of the \Nbody snapshots over the expected 
density, which is quite close to the expected unity. The three variants performed 
similarly well over most of the radial range, while there are some minor differences 
at small and large radii.
Next we compute the rate of energy diffusion as a function of particle energy, which is 
defined as the slope of the mean-square relative change of energy $(\Delta E/E)^2$ 
as a function of time, and compare it to the theoretically predicted diffusion coefficients 
(see \citet{Theuns96} for another example of such comparison). 
If the model is in a stable equilibrium, the particle energy changes only due to 
two-body relaxation, with a rate that can be computed using standard expressions 
for diffusion coefficients if one knows the distribution function 
\citep[e.g.][Eqs. 5.55, 5.125]{Merritt13}. 
The only adjustable parameter in this computation is the value of Coulomb logarithm, 
and the usual practice is to take it to be roughly $\ln N$ for the entire model, 
or $\ln M_\bullet/m_\star$ for the region around the black hole; 
we interpolated between these asymptotic regimes by taking 
$\ln\Lambda=\ln[(M_\bullet+M(E))/m_\star]$, where $M(E)$ is the mass in stars with 
binding energies higher than $E$.
Fig.~\ref{fig_d1bh001_dens_energydiff}, bottom panel, demonstrates that the energy 
diffusion rate is again very similar between different variants and close to the 
theoretical prediction (which is a non-trivial fact: in an earlier, somewhat buggy 
implementation one of the models displayed substantially larger initial diffusion 
rate before settling into a new equilibrium, while looking reasonable in other aspects). 
This plot also shows that the relaxation time, computed as the inverse of relative 
energy diffusion rate, is a few times longer than the integration time of the runs, 
therefore we do not expect the two-body effects to play a substantial role, at least 
in the initial evolution. For comparison, the dynamical time is $\sim 10^{-1}$ at 
the radius of influence and $\sim 1$ at the half-mass radius.

Finally, most interesting is the evolution of model shapes (axis ratios). 
We used an iterative procedure (see \citet{ZempGGK11} for an extended discussion) 
to determine the axes of inertia tensor of particles within a certain ellipsoidal 
radius, with the scaling of this ellipsoid's axes being iteratively updated from 
the inertia tensor until both converge to within a defined tolerance.
Plotted in Fig.~\ref{fig_d1bh001_shape} are the axis ratios ($y/x$ and $z/x$) 
as functions of time, for radii containing 2\%, 10\% and 50\% of total mass 
(not including the black hole).
Here again the three variants performed almost identically. 
Not surprisingly, the evolution was strongest in the inner parts, although 
even there the triaxiality remained substantial. 
A more detailed exploration of the effects of resolution, initial shape and other 
parameters is left for a future study.

\subsection{Cold collapse model}

Instead of creating models for a known, analytical mass distribution such as the Dehnen 
profile, we use a triaxial $N$-body model obtained by cold collapse as the target 
density profile. 
Namely, we take $N=0.5\times10^6$ equal-mass particles distributed according to 
$\rho_\mathrm{init}(r) = (2\pi\,r)^{-1}$ in the sphere of unit radius and unit mass, 
with velocities assigned from isotropic gaussian distribution with 1D velocity dispersion 
$\sigma_\mathrm{1D}\approx 0.08$, independent of radius, so that the initial virial ratio 
of the system is $3\times 10^{-2}$.
The collapse of this sphere results in a strongly triaxial system due to the development 
of radial-orbit instability \citep{PolyachenkoShukhman81, MerrittAguilar85}; 
the density in the inner part increases by a factor of few while retaining the slope,
and axis ratios vary with radius, but roughly equal to $1:0.6:0.4$.
Such cold collapse models are rather well studied \citep[e.g.][]{AguilarMerritt90, 
CannizzoHollister92, BoilyAthanassoula06};
a similar approach was used by \citet{VoglisKS02, MuzzioCW05} for the same goal of 
creating triaxial models in equilibrium, although the properties of these models are 
related to the initial conditions in a quite complicated and not always predictable way.
Unlike these studies, we follow the collapse and subsequent relaxation of the model by 
a less approximate $N$-body code. 

First we use the direct, hardware-accelerated $N$-body code $\phi$GRAPE \citep{HarfstGMSPB07} 
to follow the initial collapse and formation of triaxial bar up to $t=10$
(the peak of the density and depth of the central potential is reached around $t=1$);
at this first stage we set a very small softening length $\epsilon=10^{-4}$ (to prevent 
the formation of binaries), which is nevertheless smaller than the mean interparticle 
separation at the peak of collapse. The energy is conserved to $10^{-4}$ during this 
first stage. 
Then we followed the evolution for another 100 time units by the efficient tree-code 
\texttt{gyrfalcON} \citep{Dehnen00}, with $\epsilon=10^{-3}$, to create a well-mixed system 
(the traces of initial cold streams in phase space disappeared); the energy was conserved to 
the relative accuracy of $3\times10^{-4}$. 
The density profile did not change appreciably at this stage, apart from the expansion of 
outermost layers composed of unbound particles and some flattening at small radii.
This flattening is not unexpected, since the relaxation time in the inner parts of the 
model is quite short: the locally evaluated 
$T_\mathrm{rel}=0.34\sigma^3(r)/[m_\star\,\rho(r)\,\ln\Lambda]$ \citep{BT} 
is shorter than 100 time units for $r<0.04$, while the dynamical time at this radius 
is $\sim 0.1$. 
The density profile can be roughly described by a $\gamma=0$ Dehnen model with the scale 
radius $r_0=0.06$, but with moderate deviations and a non-constant axis ratio.
Our chosen value for the softening length is just slightly larger than the optimal value 
for a $\gamma=0$ Dehnen sphere, estimated in \citet{AthanassoulaFLB00} to be $\sim 0.01$ 
at this number of particles and for the scale radius of unity.
At the end of the second stage, we eliminated the unbound particles (which comprised 12\% 
of mass) and rotated the snapshot to be aligned with principal axes. 
In addition, the model demonstrated a slow figure rotation, similar to what has been found 
in analogous experiments \citep[e.g.][]{AquilanoMNZ07, MuzzioNZ09}, which we neutralized 
by flipping the positions and velocities of a fraction of particles about one or more 
principal planes in such a way as to make the total linear and angular momentum as small 
as possible. 
Thus we obtained the initial model for the third stage, which is expected to be an 
equilibrium configuration.

A stacked-up combination of last five snapshots from the second stage, taken at moments 
of time from 96 to 100 and joined to decrease shot noise, served as the input 
to the potential initialization (BSE and Spline potential expansions) used to generate three 
variants of Schwarzschild models (C, B and S), each having $10^5$ orbits and $\sim 500$ 
constraints. (Models C and S used the Spline potential expansion and model B used the BSE 
potential with the shape parameter $\alpha=1$, i.e. the Hernquist-Ostriker basis set).
The orbital structure of these models is quite rich, with roughly half of orbits being 
tubes (short- and long-axis tubes in comparable quantities) and the rest are boxes, 
of which about a third belong to various resonant families. Most of box orbits are chaotic 
while most tubes are not; the percentage of chaotic orbits depends on the number of terms 
in potential expansion (adding more high-frequency components makes more orbits chaotic), 
but in general about a half of all orbits are chaotic.

\begin{figure} 
$$\includegraphics[width=7cm]{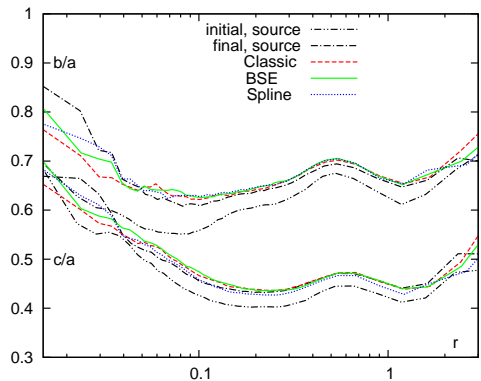} $$
\caption{
The shape of a cold collapse model (axes ratio as functions of radius). 
Black dash-double-dotted line -- the original snapshot at $t=100$ (at the end of the second stage), 
which served as the source for creating three variants of Schwarzschild models, which were 
then evolved for another 200 time units (the third stage).
Other curves are for snapshots at the end of the third stage: 
black dash-dotted line -- the source snapshot continued to evolve as it is, 
other three lines are for C, B and S variants of Schwarzschild models. \protect\\
All three variants of Schwarzschild models have attained a similar shape by the end of simulation, 
which is somewhat less flattened than the initial snapshot, but consistent with the shape 
of final snapshot evolved directly from the initial one.
}  \label{fig_collapse_shape}
\end{figure} 

The original model, composed of $4.4\times 10^5$ particles, and the \Nbody models 
generated from each variant of Schwarzschild model with the same number of particles,
were evolved for 200 time units using \texttt{gyrfalcON} with the same softening length 
as for the second stage.
Figure~\ref{fig_collapse_shape} shows that by the end of simulation, all three variants of 
Schwarzschild model have become somewhat less triaxial; the evolution of axis ratios was 
virtually the same for all models and for the original collapse simulation. 
Therefore, we have confirmed the ability of the Schwarzschild method to create strongly 
triaxial models which are as stable as the models created with other methods (in this 
example, cold collapse), which was recently called in question by \citet{ZorziMuzzio12}.

\section{Conclusions}

We reviewed the methods for studying non-spherical galactic models, analyzing properties 
and regularity of orbits and the overall structure of phase space in a given potential, 
and constructing self-consistent equilibrium models by the Schwarzschild method.

We developed a new software named \SMILE, which is intended to be a convenient 
tool for performing orbit analysis and Schwarzschild modelling for a number of standard 
potential models, as well as for an arbitrary density profile (including a discrete \Nbody 
model) approximated by a flexible potential expansion. 
The new and improved methods implemented in \SMILE include:
\begin{itemize}
\item Two general-purpose potential expansions for arbitrary density models, in which 
the angular dependence of density and potential is decomposed into spherical harmonics 
and the radial dependence of the expansion coefficients is represented either as a sum 
of basis functions (using \citet{Zhao96a} basis set, which is a generalization of the 
commonly employed \citet{HernquistOstriker92} set), or as a spline. 
Another variant of potential representation is via a set of fixed point masses, 
using a tree-code approach and a spatially-adaptive softening length.
\item An improved method of chaos detection via Lyapunov exponent $\Lambda$, which can 
distinguish between orbits having $\Lambda>0$ and those that did not demonstrate any chaotic 
behaviour over the integration time.
\item The method of generating initial conditions for a Schwarzschild model, based on sampling 
of an isotropic distribution function of an equivalent spherical model. It eliminates some 
artifacts that appear in the traditional method of assigning initial conditions on a regular grid, 
and allows to sample the phase space more uniformly. As a by-product, these spherical mass models 
may be used to efficiently analyze dynamical properties of arbitrary density profiles, including 
the ones from \Nbody snapshots.
\item Two new variants of Schwarzschild models, in which density of the target model and of the 
orbits is represented by one of the potential expansions mentioned above.
\end{itemize} 
The program comes in two versions: an integrated graphical interface allows to connect various 
related tasks and instantly analyse the results (e.g. select an orbit from the frequency map or 
the Poincar\'e section, visualize its shape and spectrum, compare orbit distributions in different 
models, etc.), while the console version is more suitable for performing scriptable tasks.
It is written in \texttt{C++} and available for download at 
\texttt{http://td.lpi.ru/\symbol{126}eugvas/smile/}.
Modular design allows to use parts of the code (e.g. orbit integration or analysis) 
in other programs, for example for analyzing orbits from an \Nbody simulation; 
also provided is integration with \texttt{NEMO} \Nbody simulation framework \citep{Teuben95}.

We have performed accuracy tests of the potential expansions and found that they may be efficiently 
used for approximating various smooth density profiles with sufficient accuracy, using a rather 
modest number of terms; the Spline potential is generally more flexible and efficient. 
For a discrete mass set, the accuracy of these smooth representations is limited by the shot noise; 
we derived the criteria for choosing an appropriate number of terms. 
Nevertheless, smooth expansions are better describing the underlying density model (sampled by 
a discrete mass set) than a direct representation of the potential using $N$ point masses, 
even if the latter uses an appropriately chosen adaptive softening.
We also tested our implementations of the classic and two new variants of Schwarzschild models 
on two test cases, and found them to perform with similar efficiency and fully capable of creating 
triaxial models in dynamic equilibrium.

\textbf{Acknowledgements:} I am grateful 
to Cs.~M{\'e}sz{\'a}ros for providing me with his linear/quadratic programming solver BPMPD, 
to David Merritt for drawing my attention to non-parametric density estimators which 
resulted in implementation the Spline potential expansion as an alternative to basis-set,
to Lia Athanassoula for fruitful discussions on the chaos-related topics, 
and to the referee, Daniel Pfenniger, for important comments concerning the new variants 
of Schwarzschild models based on expansion methods.
This work was partially supported by Russian Ministry of science and education 
(grants No.2009-1.1-126-056 and P1336), by NASA under award No. NNX10AF84G, 
by NSF under award No. AST1211602, and by CNRS which made possible the visit to 
Laboratoire d'Astrophysique de Marseille.


\appendix

\section{Potential expansions}  \label{sec_potential_expansions}

Two general-purpose potential representations used in \SMILE are the basis-set and Spline expansions. 
They share the method of dealing with angular dependence of density and potential by expanding 
it in terms of spherical harmonics $Y_l^m(\theta, \phi)$, but differ in the description of radial 
variation of expansion coefficients. In addition, the same angular expansion may also be used as 
an approximation to the scale-free (single power-law) density profile. 
The angular dependence of density and potential is given by
\newcommand{\trig}{\mathrm{trig\,}}
\begin{eqnarray*}
\rho(r,\theta,\phi) &\!=\!& \sum_{l=0}^{\lmax} \sum_{m=-l}^{l} 
  A_{lm}(r)\,\sqrt{4\pi} \tilde P_l^m(\cos\theta)\,\trig m\phi \,, \\
\Phi(r,\theta,\phi) &\!=\!& \sum_{l=0}^{\lmax} \sum_{m=-l}^{l} 
  C_{lm}(r)\,\sqrt{4\pi} \tilde P_l^m(\cos\theta)\,\trig m\phi \,, \\
\trig m\phi &\!\equiv\!& \left\{\begin{array}{rcl} 
  1 &,& m=0 \\
  \sqrt{2}\,\cos  m \phi &,& m > 0 \\
  \sqrt{2}\,\sin |m|\phi &,& m < 0 
\end{array}\right.   \nonumber
\end{eqnarray*}

Here we expanded the spherical harmonics: 
$Y_l^m(\theta, \phi) = (-1)^m\, \tilde P_l^m(\cos\theta)\,\mathbf{e}^{im\phi}$, 
where $\tilde P_l^m(x)$ are normalized associated Legendre polynomials, 
and combined the exponents to obtain real coefficients with sines and cosines.
The factors $\sqrt{2}$ for $m\ne 0$ are introduced to keep $\sum_{m=-l}^{l} A_{lm}^2$ 
rotationally invariant. 
We introduced the convention that $m<0$ corresponds to odd (sine) part of expansion.
For triaxial systems only terms with nonnegative even $l$ and $m$ are nonzero, 
however we keep the more general form of expansion, while using only the necessary number of 
nonzero terms in the actual computations (e.g.\ only $l=0$ terms in the axisymmetric case).

\subsection{BSE for a scale-free potential}  \label{sec_app_scalefree}

To start with, consider a simpler case of a scale-free potential, where 
$A_{lm}(r) = A_{lm} r^{-\gamma}$, $C_{lm}(r) = C_{lm} r^{2-\gamma}$. 
The expansion coefficients for density and potential are related by the formula
\begin{equation}  \nonumber
C_{lm} = \frac{4\pi}{(l+3-\gamma)(2-l-\gamma)}\, A_{lm} \,.
\end{equation}

We note that this representation is much more accurate for a given $\lmax$ than 
cosine expansion used in \cite{Terzic02}.

\subsection{BSE for a generic triaxial potential}  \label{sec_app_bse}

Next we come to representation of generic 3-dimensional potential-density pair in terms of 
basis function series, defining
\begin{equation}  \nonumber
A_{lm}(r) = \sum_{n=0}^{\nmax} A_{nlm} \,\rho_{nl}(r) \;,\;
C_{lm}(r) = \sum_{n=0}^{\nmax} A_{nlm} \,\Phi_{nl}(r) \,.
\end{equation}

We use the $\alpha-$model basis set described by \citet{Zhao96a}. 
In the zeroth order it represents the following potential-density pair:
\begin{equation}  \nonumber 
\Phi_{00} = -\frac{1}{(1+r^{1/\alpha})^{\alpha}} \,,\quad
\rho_{00} = \frac{1+\alpha}{4\pi\alpha} \frac{1}{r^{2-1/\alpha}\,(1+r^{1/\alpha})^{2+\alpha} } \,.
\end{equation}

Setting $\alpha=1/2$ gives the Plummer potential and associated \citet{CluttonBrock73} basis set;
value of $\alpha=1$ corresponds to the \citet{Hernquist90} model and the \citet{HernquistOstriker92} 
basis set. The higher-order terms are given by
\begin{eqnarray*}
\Phi_{nl} &=& -\frac{r^l}{(1+r^{1/\alpha})^{(2l+1)\alpha}} \,G_n^{w}(\xi) \,,  \label{eq_basis_phi} \\
\rho_{nl} &=& \frac{K_{nl}}{2\pi} \frac{r^{l-2+1/\alpha}}{(1+r^{1/\alpha})^{(2l+1)\alpha+2}} 
  \, G_n^{w}(\xi) \,, \label{eq_basis_rho} 
\end{eqnarray*}
\begin{equation}  \nonumber
  K_{nl} \equiv \frac{4(n+w)^2-1}{8\alpha^2} ,\;
  w \equiv (2l+1)\alpha+1/2 \;,\;
  \xi \equiv \frac{r^{1/\alpha}-1}{r^{1/\alpha}+1} ,
\end{equation}
where $G_n^{w}(\xi)$ are Gegenbauer (ultraspherical) polynomials. 

\subsection{Spline spherical-harmonic expansion}  \label{sec_app_spline}

Another option for approximating an arbitrary potential is to represent its angular part in 
spherical harmonics, but retain an explicit radial dependence of the expansion coefficients 
$C_{lm}(r)$, with forces and density given by derivatives of these functions. 
Accordingly, $C_{lm}(r)$ is represented as a cubic spline interpolating between a grid 
of nodes. A straightforward approach faces some difficulties: it is hard to accomodate 
a large dynamic range in radii and, more importantly, ensure that even the second 
derivative of $C_{lm}(r)$ is reasonably close to that of the underlying model.

We use the following modification to the base scheme: first, the lowest-order 
term $C_{00}$ is represented as a double logarithmic function:
\begin{equation}  \nonumber
\tilde C_{00}(\xi\equiv\log r) \equiv -\log(C_{00}(0)^{-1}-C_{00}(r)^{-1}) \,.
\end{equation}

The advantage of such scaling is that it can accomodate a power-law behaviour at
small $r$: if $C_{00}(r) \approx C_{00}(0)\times(1-Kr^{2-\gamma})$, then 
$\tilde C_{00}(\xi) \approx \log(-C_{00}(0)/K) - (2-\gamma)\xi$ 
as $\xi\to -\infty$. 
This modified function is approximated by a cubic spline in $\log r$ and 
extrapolated at both small and large $r$: for $\log r\to -\infty$ we use the
linear extrapolation described above, which corresponds to powel-law density 
behaviour at $r\to 0$. For $r\to \infty$ the extrapolation is mimicking a 
power-law density profile $\rho \propto r^{-\gamma_\mathrm{out}}$ with 
$\gamma_\mathrm{out}>3$; in this case 
\begin{equation}  \nonumber
C_{00}(r) = C_{00}(r_\mathrm{out})\, (r_\mathrm{out}/r)\,
[1-\mu\{(r_\mathrm{out}/r)^{\gamma_\mathrm{out}-3} -1\}] \,,
\end{equation}
where $\gamma_\mathrm{out}$ is calculated from the last three grid nodes and 
$\mu$ -- from the last two.

Higher-order coefficients cannot be represented in such a logarithmic way 
because they need not have the same sign over all $r$; instead, they are 
normalized by $C_{00}(r)$ and interpolated in $\zeta \equiv\log(1+r)$: 
the spline is constructed for 
\begin{equation}  \nonumber
\tilde C_{lm}(\zeta) \equiv C_{lm}(r)/C_{00}(r) \,.
\end{equation}

The evaluation of ``real'' (not tilded) coefficients and their derivatives is 
just a number of simple algebraic transformations of the spline values and 
derivatives up to second order, which are trivially obtained for a cubic spline.
Inside the first and beyond the last grid nodes, coefficients are extrapolated 
as power-law functions of radius.

The radial grid for representing the coefficients is taken to be exponentially spaced:
$r_k = r_1\times[\exp(Z\,k)-1]/[\exp(Z)-1]$, where $r_1$ is the radius of the first node, 
and $Z$ is assigned a value which gives some predefined radius $r_G$ for the last node
$\ngrid$. Typically, it makes sense that $r_1$ encloses some small fraction of mass 
($\delta \sim 10^{-3}$ or less), and $r_G$ encloses $1-\delta$ of total mass; 
logarithmic scaling makes a very good approximation with as small as $\ngrid=10-20$ 
nodes even for $r_G/r_1 \gtrsim 10^4$.

If the potential is initialized from a set of $n_\mathrm{pt}$ discrete point masses, 
the splines are constructed using an adjustable amount of smoothing, according to the 
following procedure. First, the inner and outer grid radii $r_1, r_G$ are assigned so 
that there are only a few particles inside $r_1$ or beyond $r_G$, 
and the radial grid is constructed using exponentially spaced nodes.
Then the expansion coefficients $C_{lm}(r_i)$ are computed at each particle's radius 
$r_i$, $i=1..n_\mathrm{pt}$. and the penalized least-square fitting method is used 
to construct a smoothing spline $\tilde C_{lm}(x)$ which minimizes the following functional:
\begin{equation}  \nonumber
\mathcal{L} \equiv \sum_{i=1}^{n_\mathrm{pt}} \left\{C_{lm,i}-\tilde C_{lm}(x_i)\right\}^2 
  + \lambda \int \left\{ \tilde C_{lm}''(x)\right\}^2\,dx \,.
\end{equation}
Here $x$ are the scaled radial variables ($\xi$ or $\zeta$), and $\lambda$ is 
the smoothing parameter. 
As discussed in \citet{MerrittTremblay94}, increasing this parameter to infinity is equivalent 
to fitting the data with a pre-specified functional form of regression; in general, unlike 
kernel-based smoothing approaches, minimization of $\mathcal{L}$ tends to produce power-law 
regressions, in this case the $l=0$ term corresponds to the Hernquist profile
(for comparison, smoothing the mass profile as described in section~\ref{sec_schw_spherical} 
would produce a two-power-law model $\rho(r) \propto r^{-\gamma}(r_0+r^{3-\gamma})^{-2}$ 
in the limit of infinite $\lambda$).
The value of the smoothing parameter $\lambda$ may be chosen at will; while there are standard 
techniques such as generalized cross-validation method \citep{Wahba90}, they tend to produce 
very little smoothing if $n_\mathrm{pt}\gg \ngrid$. A practical recipe implemented in \SMILE 
uses the Akaike information criterion \citep[AIC, e.g.][]{BurnhamAnderson02} in such 
a way that the difference $\Delta$AIC between the fit with no smoothing and the fit 
taken to be ``optimal'' does not exceed some predefined (adjustable) value. 
Typically this results in little extra smoothing for low $l,m$ where coefficients 
do indeed contain useful data, and a substantial reduction of noise for higher $l,m$.

\label{lastpage}

\end{document}